\documentclass[%
reprint,
superscriptaddress,
 amsmath,amssymb,
 aps,
prb,
]{revtex4-2}

\usepackage{graphicx}
\usepackage{dcolumn}
\usepackage{bm}
\usepackage{hyperref}
\hypersetup{colorlinks=true,citecolor={blue},linkcolor={blue},urlcolor={blue}}
\usepackage{xcolor}

\usepackage{soul}
\begin{document}

\preprint{APS/123-QED}

\title{Optically controlled polariton condensate molecules}

\author{E.$\,$D. Cherotchenko}
\affiliation{Ioffe Physical-Technical Institute of the Russian Academy of Sciences, 194021 St. Petersburg, Russia}

\author{H. Sigurdsson}
\affiliation{School of Physics and Astronomy, University of Southampton, Southampton SO17 1BJ, United Kingdom}

\author{A. Askitopoulos}
\affiliation{Skolkovo Institute of Science and Technology, Bolshoy Boulevard 30, bld. 1, Moscow, 121205, Russia}

\author{A.$\,$V. Nalitov}
\affiliation{ITMO University, St. Petersburg 197101, Russia}
\affiliation{Faculty of Science and Engineering, University of Wolverhampton, Wulfruna St, Wolverhampton WV1 1LY, UK}
\affiliation{Institut Pascal, PHOTON-N2, Universit\'e Clermont Auvergne}

\date{\today}

\begin{abstract}
A condensed matter platform for analogue simulation of complex two-dimensional molecular bonding configurations, based on optically trapped exciton-polariton condensates is proposed. The stable occupation of polariton condensates in the excited states of their optically configurable potential traps permits emulation of excited atomic orbitals. A classical mean field model describing the dissipative coupling mechanism between $p$-orbital condensates is derived, identifying lowest threshold condensation solutions as a function of trap parameters corresponding to bound and antibound $\pi$ and $\sigma$ bonding configurations, similar to those in quantum chemistry.
\end{abstract}

\maketitle


\section{Introduction}
\label{sec:intro}

The idea of simulating complex molecular systems by mapping them onto experimentally accessible and transparent platforms maintains growing interest due to technological advancements in designing quantum simulators~\cite{Georgescu_RMP2014}. In this paradigm, quantum computers~\cite{Aspuru_Science2005} step in for CMOS computer technologies to calculate and predict meaningful quantities in quantum chemistry. Amazing progress in quantum simulation for chemistry problems has been made in the last several years using superconducting qubits~\cite{Malley_PRX2016, Kandala_Nature2017, Arute_Science2020} and integrated photonic circuitry~\cite{Lanyon_NatChem2010, Peruzzo_NatComm2014}. On the other hand, analogue quantum simulators, operating on the principle that the natural dynamics of a specially designed system will emulate a target quantum object, have been implemented using cold atom ensembles~\cite{Luhmann_PRX2015, Arguello_Nature2019} with good success. Here we propose such a platform based on exciton-polariton condensates, macroscopic quantum objects, for analogue simulation of two-dimensional molecular bonding between atoms.

Exciton-polariton condensates in planar microcavities can be confined in etched cavity pillars and mesas, forming a discrete photonic spectrum similar to atomic orbitals~\cite{kaitouni_engineering_2006, Bajoni_PRL2008, Ferrier_PRL2011}.
When combined, adjacent micropillars become coupled due to the overlapping orbitals forming new macroscopic states~\cite{Galbiati_PRL2012, Sala_PRX2015, Mangussi_IOP2020}. Exciton-polaritons may also be confined optically, by using non-resonant lasers which excite a background exciton gas that repulsively interacts with polaritons~\cite{Askitopoulos_PRB2013, Cristofolini_PRL2013}.
Stimulated bosonic scattering from the surrounding exciton gas then provides gain to the confined condensate, compensating for photonic losses due to the finite microcavity quality factor. In optical traps, polariton condensates demonstrate rich phenomenology, including spin bifurcations~\cite{Ohadi_PRL2017}, quantum chaos~\cite{Gao_Nature2015}, vortices~\cite{Gao_PRL2018, Ma_NatCom2020}, spin hysteresis~\cite{Pickup_PRL2018}, and multilevel synchronization between traps~\cite{tpfer2020lotkavolterra}. 

The advantage of using polariton condesates as macroscopic emulators of quantum objects is the easy readout of the polariton state through the emitted cavity light. It permits both integrated and time-resolved measurement of the energy resolved real-space and momentum-space condensate density, phase, spin (polarization), and correlations between selected condensates~\cite{topfer_engineering_2020}. In contrast, high resolution time-of-flight experiments using ultracold atoms can only indirectly measure correlations and phase of the atomic ensembles by relying on symmetry arguments and global properties of the atoms reciprocal density distribution~\cite{Luhmann_PRX2015}. 

Of interest, compared to other methods of trapping polariton condensates~\cite{Maragkou_PRB2010, Galbiati_PRL2012}, optical traps result in stable occupation of the excited states~\cite{Manni_PRL2011, Tosi_NatPhys2012, Dreismann_PNAS2014, Askitopoulos_PRB2015, Sun_PRB2018, Xu_OptExp2019, tpfer2020lotkavolterra} due to the optical gain provided by the laser. That is, more energetic states penetrate further into the laser-induced potential, experiencing enhanced scattering into the condensate, and remain indefinitely stable as long as the laser is present. This is in contrast to equilibrium condensates which instead relax quickly to the ground state. Moreover, optically induced potentials benefit from being completely reconfigurable by simply reprogramming the shape of the external nonresonant laser light using liquid crystal spatial light modulators, avoiding the costs of fabricating new patterned cavities from scratch. These two properties of optical traps, stable condensation into exited orbitals, and flexibility in potential engineering, underpin their appropriateness for possible analogue simulation of molecular bonding.

Today, coupled polariton condensate have been proposed to simulate the energy minima of the XY spin system~\cite{berloff_realizing_2016} and recent works have started to address their potential to simulate molecular bonding~\cite{alex2020artificial}. There, each condensate is cylindrically symmetric and radially emits polaritons equally in all directions. Thus the coupling strength between adjacent condensates depends only on the spatial separation distance and occupation difference between condensates. However, if individual condensates occupy more complex orbitals the in-plane emission of polaritons from the condensate becomes anisotropic and the coupling strength will become dependent on the relative angle between neighbors. To our knowledge, this degree of complexity has not been considered before in driven-dissipative systems where the coupling becomes defined by a new observable; the internal angle of the condensate ``atom''.

In this work we show that dissipative coupling of optically trapped and spatially separated polariton condensates demonstrates unique features even in the linear regime close to the condensation threshold. Namely, self-ordering of the macroscopic polariton orbitals orientation to optimize the system gain. The linear specifics of this mechanism are pronounced in coupling of degenerate excited trapped states.
In particular, we consider the coupling of the first excited doublet, sharing similarity to atomic $p$-orbitals. Coupled trapped condensates demonstrate mutual alignment of dipole spatial wavefunction profiles, similarly to $\pi$ and $\sigma$ molecular bond types.
The particular type of bond alignment due to the dissipative mechanism depends on the separation distance and the angle between traps.
This observation paves the way towards optically controlled emulation of coupling and hybridization of electronic orbitals in molecules.

We first develop the complex delta shell potential model of a single trap and compute its eigenstates in Section \ref{sec:level1}.
Treating coupled traps within the perturbation theory, we then compute the spectrum of synchronized states in Section \ref{sec:coupled}.
We calculate the dependence of interaction strength on the distance between the traps and identify four possible types of coupling and alignment of dipole $p$-state condensates in adjacent traps in Subsection \ref{subsec:long} and \ref{subsec:general} respectively.
Polariton spectrum and condensation in the presence of considered interaction are analysed in the cases of two traps in Section \ref{sec:two}.
Polariton condensation into aligned dipole states is further confirmed with numerical simulation of the complex Gross-Pitaevskii equation.

\section{Single trap}
\label{sec:level1}

Dynamics of polariton condensates interacting with spatially inhomogeneous exciton reservoirs is governed by the generalized (driven-dissipative) Gross-Pitaevskii equation, coupled to the semiclassical rate equation for the reservoir density~\cite{Wouters2007a}:
\begin{align}
i\hbar {\partial\Psi\over\partial t} & = \left[ - { \hbar^2 \Delta \over 2 m^*} + {\alpha + i \beta \over 2}N + \alpha_1 \vert \Psi \vert^2 - i \hbar {\Gamma \over 2}  \right] \Psi, \label{gpe}\\
{\partial N \over \partial t} & = P(r) - (\beta \vert \Psi \vert^2 + \gamma) N. \label{res}
\end{align}
Here, $\Delta$ denotes the two-dimensional Laplacian operator, $\Psi$ and $N$ are the condensate wavefunction and the reservoir density, $m^*$ is the  polariton effective mass, $\alpha$ and $\alpha_1$ are the interaction constants describing the polariton repulsion off the exciton density and polariton-polariton repulsion, $\beta$ governs the stimulated scattering from the reservoir into the condensate, $\Gamma$ and $\gamma$ are the polariton and exciton decay rates, and $P(r)$ is the cylindrically symmetric reservoir pumping rate. We will focus on an annular shaped laser pumping profile, similar to those used in~\cite{Manni_PRL2011, Askitopoulos_PRB2013, Cristofolini_PRL2013, Dreismann_PNAS2014, Askitopoulos_PRB2015, Sun_PRB2018, Harrison_PRB2020, tpfer2020lotkavolterra}. The laser pump is approximated with a delta shell ring shape $P(r) = P_0 \delta(r-R)/R$, where $r =\sqrt{x^2 + y^2}$ is the radial coordinate, $R$ is the trap radius, and $P_0$ represents the pumping power.

Close to the condensation threshold (i.e., normal-to-condensed state phase transition) one may put $\vert \Psi \vert^2 \simeq 0$, neglecting polariton-polariton interactions and reservoir depletion in Eqs.~\eqref{gpe} and~\eqref{res}. 
This approximation allows establishing the hierarchy of fastest exponential decay (or growth) rates of polariton modes in the system. At threshold, scattering from the reservoir starts exceeding the polariton decay rate, corresponding to a mode with the fastest population growth rate. At higher laser power, beyond the condensation threshold, the condensate state is typically governed by this fastest growing mode which saturates the reservoir and stabilizes. Polariton interactions, however, can destabilize this state when the mean field energy becomes comparable with other characteristic energy scales~\cite{Aleiner2012}. We will not consider this regime and focus instead on hierarchy of the fastest growing modes around threshold.

The solutions corresponding to the linear limit of Eq.~\eqref{gpe} are stationary, allowing us to put the left-hand part of Eq. \eqref{res} to zero and express the reservoir density as $N = P(r)/\gamma$.
The reservoir effect on the condensate then reduces to a complex potential given by
\begin{equation}
    V(r) = {V_0}\frac{\delta(r-R)}{R},
\end{equation}
where the potential strength $V_0 = (\alpha + i \beta) P_0 /(2 \gamma R)$. The generalized Gross-Pitaevskii equation then becomes a dissipative (non-Hermitian) Schr\"{o}dinger equation for the polariton wave function, which in polar coordinates reads:
\begin{equation} \label{Shroed}
   \left[ - \frac{\hbar^2}{2m^*}\left(\frac{1}{r}\frac{\partial}{\partial r}r\frac{\partial}{\partial r}+\frac{1}{r^2}\frac{\partial}{\partial \varphi^2}\right )+V(r) - i\hbar \frac{\Gamma}{2}\right]\Psi = E\Psi.
\end{equation}
Where $E$ is the complex valued eigenenergy.

Rotational symmetry of the problem allows us to separate the angular and radial coordinates and express the wavefunction in the form $\Psi = \psi(r)\Phi(\varphi)$, where $\Phi(\varphi) = e^{im\varphi}$.
Substituting into Eq.~\eqref{Shroed} results in an equation for the radial part:
\begin{equation}
\label{Bess1}
   \left[ -\frac{\hbar^2}{2m^*}\left(\frac{1}{r}\frac{d}{dr}r\frac{d}{dr}-\frac{m^2}{r^2}\right )+ V(r)-E-i\hbar\frac{\Gamma}{2}  \right]\psi = 0.
\end{equation}
As we are considering the delta-functional potential, inside and outside the trap the equation will take the form:
\begin{equation}
r^2\frac{d^2 \psi}{dr^2}+r\frac{d \psi}{dr}+\left[ 2m^* \left(E+i\hbar\frac{\Gamma}{2} \right) \frac{r^2}{\hbar^2}-m^2\right] \psi=0.
\end{equation}
With the substitution $\rho(r) = \sqrt{2m^* \left(E+i\hbar\Gamma/2 \right) }r / \hbar$ (the square root corresponds to the principle value) the equation turns into the canonical Bessel equation.
In the region $r<R$ the wavefunction then reads $\psi = AJ_m(\rho)$, where $J_m(\rho)$ is the Bessel function of the first kind and we have taken into account the divergence of the Bessel function of second kind $Y_m(\rho)$ at the origin. 
For $r>R$ the solution can be expressed as $\psi = BH^{(1)}_m(\rho)$, where Hankel function of the first kind is chosen to converge at $r \rightarrow \infty$, keeping in mind that $\mathrm{Im}\left\lbrace E + i \Gamma /2 \right\rbrace>0$ and $\mathrm{Re}\left\lbrace E + i \Gamma /2 \right\rbrace > 0$.
In order to match the solutions at the point $r=R$ one needs to satisfy,
\begin{equation}
    A J_m(\rho_R) = B H_m^{(1)}(\rho_R)
\end{equation}
and then should integrate Eq.~\eqref{Bess1} over the small region $r \in \{R- \delta R,R+\delta R\}$, that results in the following condition:
\begin{equation}
\left.\frac{d\psi}{dr}\right|_{R-\delta R}^{R+\delta R}=\frac{c}{R}\psi(R),
\end{equation}
amwhere $c=2m^*V_0/\hbar^2$.
One can derive for $\delta R \rightarrow 0$:
\begin{align} \label{eq_dpsi1}
\left.\frac{d\psi}{dr}\right|_{R-\delta R}=&{A\rho_R \over R}\left(-J_{m+1}(\rho_R)+\frac{m}{\rho_R}J_m(\rho_R)\right), \\
\left.\frac{d\psi}{dr}\right|_{R+\delta R}=&{B\rho_R \over R}\left(-H_{m+1}^{(1)}(\rho_R)+\frac{m}{\rho_R}H_m^{(1)}(\rho_R)\right), \label{eq_dpsi2}
\end{align}
where $\rho_R = \rho(R)$.
Equations~\eqref{eq_dpsi1}-\eqref{eq_dpsi2} yield the traps resonance condition, given by the transcendental complex equation
\begin{equation} \label{eq_resonance}
\frac{J_m(\rho_R)\left( mH_m^{(1)}(\rho_R)-\rho_RH_{m+1}^{(1)}(\rho_R)\right)}{H_m^{(1)}(\rho_R)\left( mJ_m(\rho_R)-\rho_R J_{m+1}(\rho_R)+cJ_m(\rho_R)\right)}=1.
\end{equation}
\begin{figure}
\center 
\includegraphics[width=\linewidth]{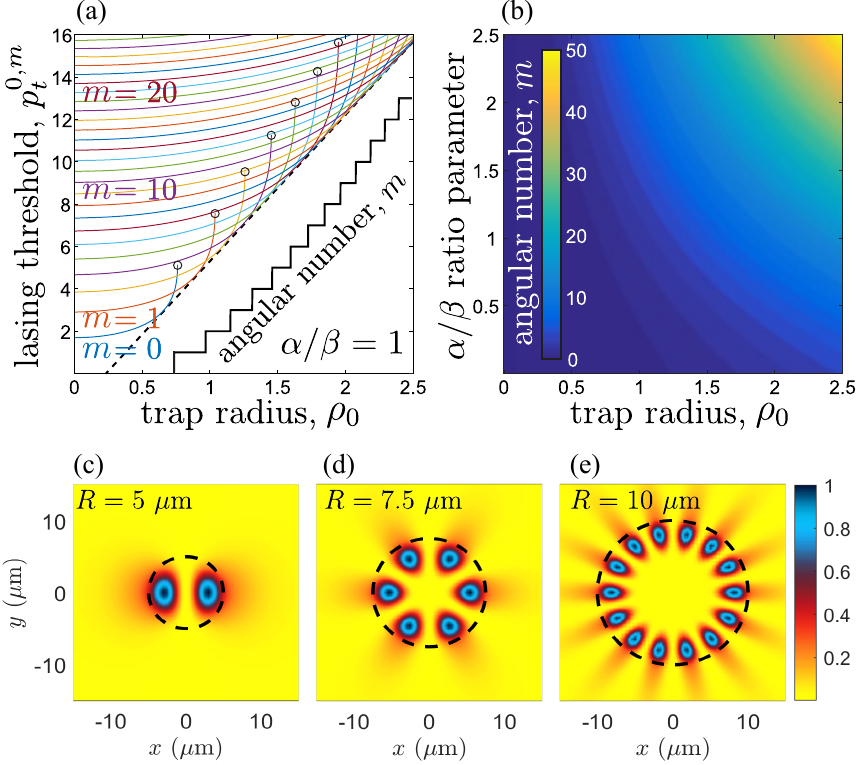}
\caption{(a) Condensation (lasing) threshold as functions of the trap radius for $\alpha/\beta = 1$.
Each state is only sustainable below a certain critical radius (black circles). The threshold state angular number $m$ switches in a ladder-like dependence with trap radius (black solid line). The minimal threshold for wide traps ($\rho\gg1$) has a linear characteristic dependence on $\rho$, shown with a dashed line. (b) Angular number of the state, at which condensation occurs. (c-e) Simulated lowest threshold steady state solutions $|\Psi(\mathbf{r})|^2$ of Eqs.~\eqref{gpe}-\eqref{res} for a single trap of different radii $R = 5, \ 7.5, \ 10$ $\mu$m, obtained for powers $P_0/R = 25, \ 40, \ 55$ $\mu$m$^{-2}$ ps$^{-1}$ respectively. The trap ridge is given by the dashed black line.
\label{fig1}}
\end{figure}

The condensation threshold condition can be defined as the moment when all eigenvalues satisfy $\text{Im}\{E \}<0$ (lossy modes) except for one which satisfies $\mathrm{Im}\left\lbrace E \right\rbrace = 0$. Any state defined by its angular and the radial quantum numbers $m$ and $n$ respectively can then be associated with the so-called threshold pump power $P_t^{n,m}$ when the above condition is satisfied. This quantity may be nondimensionalized for convenience:
\begin{equation} \label{eq.Pth}
p_t^{n,m} = {(\alpha + i \beta) m^* \over \gamma \hbar^2} P_t^{n,m}.
\end{equation}
The dimensionless trap radius, in turn, reads $\rho_0 = R \sqrt{m^* \Gamma/\hbar}$. We focus on solutions of the resonance condition \eqref{eq_resonance} satisfying $\mathrm{Im}\left\lbrace E \right\rbrace =0$. Physically, these solutions would appear first in the experimentally measured cavity photoluminescence intensity with increasing pump power. We are only interested in the states $n=0$ as these possess the lowest threshold powers among the states with a given angular number $m$~\footnote{Condensation into the state $n=1$ was observed in Ref.~\cite{Dreismann_PNAS2014} when the pumping had the shape of two concentric rings}. This is in agreement with previous experimental observations \cite{Manni_PRL2011,Sun_PRB2018,Dreismann_PNAS2014}.

The solutions are summarized in Fig.~\ref{fig1}(a).
For small traps, as expected, the condensate forms at the state $m=0$.
However, for larger traps the ground state surpasses the first excited doublet $m=\pm 1$ in threshold pumping power. Consequent cascade of successive switching of states with increasing angular numbers results in a dependence of the minimal threshold, resembling a linear one at large scale (see dashed line in Fig.~\ref{fig1}(a)). In Fig.~\ref{fig1}(b) we plot an exhaustive phase diagram of the lowest threshold angular number as a function of two dimensionless parameters, the interaction parameter ratio $\alpha/\beta$ and the trap radius $\rho_0$. In Figs.~\ref{fig1}(c)-\ref{fig1}(e) we show real-space densities $|\Psi(\mathbf{r})|^2$ of the lowest threshold stationary solutions from Gross-Pitaevskii simulations [Eqs.~\eqref{gpe} and~\eqref{res}] using a pump profile corresponding to current experimental capabilities,
\begin{equation}
    P(r) = \frac{P_0}{R} \frac{L_0^4}{(r^2 - R^2)^2 + L_0^4}.
\end{equation}
Here, $L_0$ denotes the finite thickness of the reservoir induced trap walls. In all simulations throughout the paper we have chosen parameters similar to previous works: $m^* = 0.3$ meV ps$^2$ $\mu$m$^{-2}$, $\alpha_1 = 3.3$ $\mu$eV $\mu$m$^{2}$, $\alpha = 2\alpha_1$, $\beta = 1.4 \alpha_1$, $\gamma = \Gamma = 0.2$ ps$^{-1}$, $L_0 = 3$ $\mu$m.

\section{Coupling between traps}
\label{sec:coupled}
We will now formulate our theory describing the anisotropic nature of the ballistic coupling between spatially separated polariton condensates, each populating an excited state in its respective trap. It should be underlined that the mechanism of coupling and synchronization in optically trapped polariton condensates~\cite{Harrison_PRB2020} can be drastically different from its counterpart in microcavity pillar structures~\cite{Galbiati_PRL2012}. In the latter case, polariton tunneling between overlapping orbitals is responsible for their coupling, realizing the familiar bosonic Josephson junction. In the former case, however, condensates excited by the nonresonant laser beams can stabilize and phase-lock over long separation distances due to so-called dissipative (or radiative) coupling mechanism~\cite{Aleiner2012, Ohadi_PRX2016}. Dissipative coupling refers to a coupling mechanism which changes the imaginary part of the system complex energies. In other words, it is the imaginary (non-Hermitian) part from the overlap integrals between two wavefunctions and the potential which they interfere with $\langle \psi_1 | V(\mathbf{r})| \psi_2 \rangle$. It is responsible for emergence of the weak lasing regime~\cite{Aleiner2012, Zhang2015}, limit cycles~\cite{Nalitov_PRA2019, Topfer_ComPhys2020}, and chaos~\cite{Ruiz_PRB2020}. As discussed in the context of Eq.~\eqref{eq.Pth}, the imaginary part of a mode's complex energy corresponds to its gain (like optical gain in lasers), and the mode with the highest imaginary part will be occupied by the condensate. 

\begin{figure}
\centering 
\includegraphics[width=0.9\linewidth]{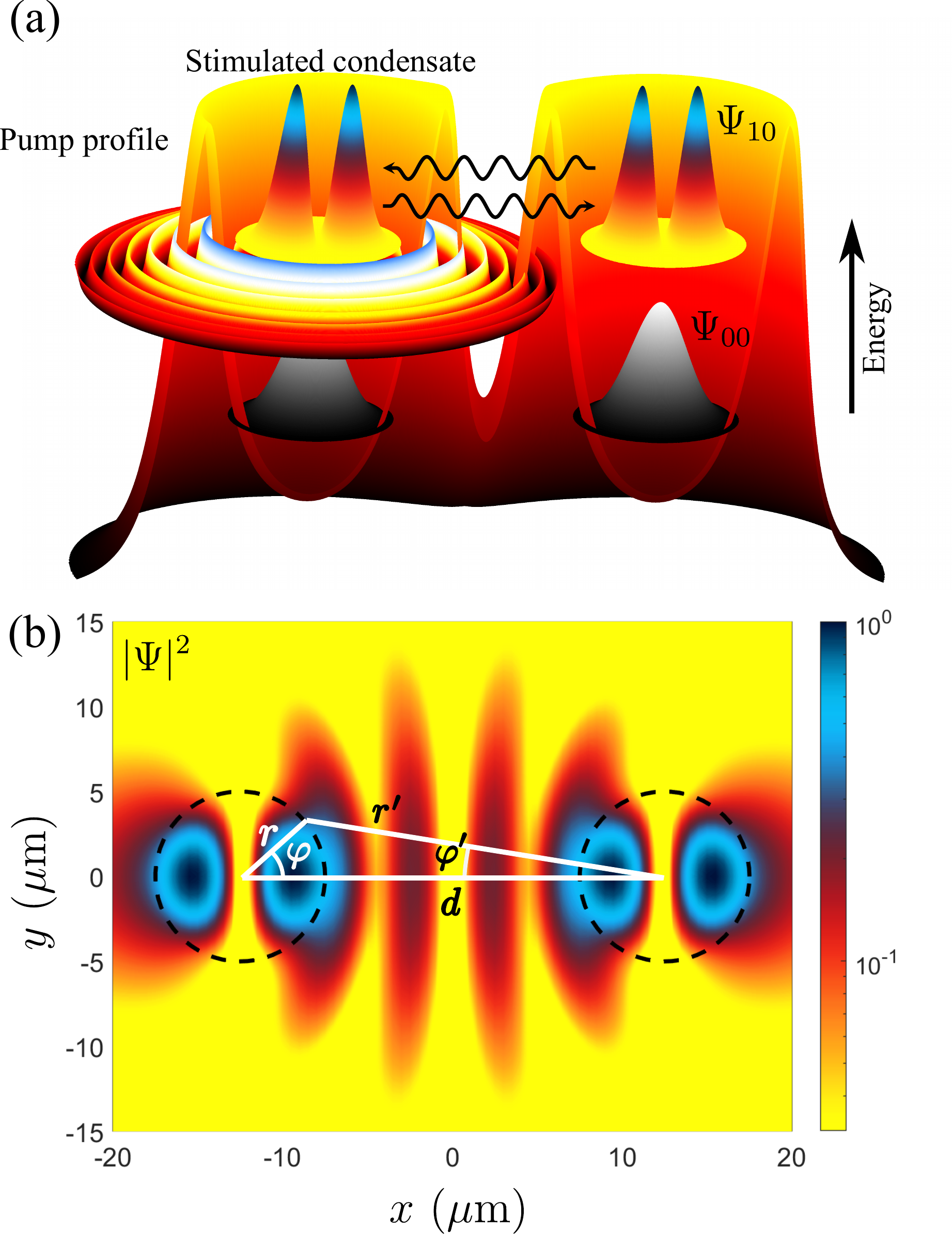}
\caption{(a) Schematic of two circular pump induced potential traps containing the first two Laguerre-Gaussian modes, the ground state $\Psi_{00}$ and the degenrate first excited state $\Psi_{10}$. Finite tunneling probability causes polariton waves to leak radially away from the trap an interfere with any neighbouring condensates. (b) Steady state simulation for $d = 25$ $\mu$m showing $|\Psi(\mathbf{r})|^2$ with clear interference fringes between two phase locked dipole-shaped condensates analogous to the chemical $\sigma$-bonding configuration. Notation of the manuscript is overlaid for clarity. Black dashed circles denote the ridges of the potential traps.}
\label{fig2}
\end{figure}

Furthermore, depending on the shape of the laser, ballistic (high kinetic energy) emission of polaritons from the condensate can become pronounced [see e.g. concentric rings in Fig.~\ref{fig2}(a)] which leads to strong matter-wave interference between neighbors. In contrast to conventional Josephson (evanescent) coupling, the range of this ballistic (and dissipative) coupling mechanism is only limited by the polariton mean free path and has been reported to synchronize condensates over hundred microns~\cite{Topfer_ComPhys2020}, about $50\times$ longer than the condensate full-width-half-maximum.

Assuming that the threshold is reached for the first excited mode doublet $m = \pm 1$, like shown schematically in Fig.~\ref{fig2}, the general form of the condensate wavefunction outside the trap reads,
\begin{equation} \label{eq.single}
    \Psi = \left( c_+ e^{i \varphi} + c_-e^{-i \varphi} \right) H_1^{(1)}(\rho).
\end{equation}
It is worth noting that these states have been realized several times in experiment~\cite{Cerna_PRB2009, Askitopoulos_PRB2015, Gao_PRL2018, askitopoulos_all-optical_2018, Ma_NatCom2020} but have lacked the theory explaining their coupling in extended systems. Here, the two complex coefficients $c_\pm$ are related to the pseudovector $\mathbf{s} = (s_x,s_y,s_z)^T$, which defines its state up to a common gauge invariant phase:
\begin{align} \notag
    s_x + i s_y  &= c_+^* c_-, \\
     s_z  &= (\vert c_+ \vert^2 - \vert c_- \vert^2)/2. \label{eq.pseudo}
\end{align}
In this notation, $s_z = \pm1$ corresponds to the two opposite circulating vortex states whereas $s_x = \pm1$ and $s_y = \pm1$ correspond to a dipole state orientated horizontally, vertically, diagonally, and antidiagonally respectively, analogous to the Stokes components of light.

For a fixed value of the condensate energy $E$ we may introduce the characteristic wavevector of the condensate outside the trap $k = \sqrt{2m^*\left( E + i\hbar \Gamma/2 \right)}/\hbar$.
A pair of coupled condensates in optically induced traps, separated by the distance $d > 2 R$, measured from the center of each trap, are then given by a common wavefunction defined by four coefficients $c_{1\pm}$, $c_{2\pm}$:
\begin{align} \notag
    \Psi(r,\varphi) = \left(c_{1+} e^{i \varphi} + c_{1-}e^{-i \varphi} \right) H_1^{(1)}(kr)  \\  \label{eq.wavef}
    +\left(c_{2+} e^{i(\pi- \varphi^\prime)} + c_{2-}e^{-i (\pi-\varphi^\prime}) \right) H_1^{(1)}(kr^\prime),
\end{align}
where
\begin{align} \label{eq_rprime}
&r^\prime = \sqrt{r^2 + d^2 - 2 rd \cos(\varphi)}, \\
&\sin(\varphi^\prime) = \sin(\varphi) r/r^\prime, \label{eq_phiprime}
\end{align}
from the law of cosines and sines respectively.

In order to investigate the growth rates of coupled states stemming from their interference we directly calculate the overlap integrals between the two wavefunctions and their respective potentials,
\begin{equation}
I = I^{(1)} + I^{(2)} = \int \Psi^* \left[ \delta(r-R) + \delta(r' - R) \right] \Psi \, d\mathbf{r}.
\end{equation}
The common (superposed) wavefunction overlap with the potential is composed of two terms, corresponding to the two excitonic traps. Let us find them separately and begin with the overlap with the first trap $I^{(1)} = \int_0^{2\pi} \left\vert \Psi(R,\varphi) \right\vert^2 d \varphi$, which, in turn, may be decomposed into the sum of overlaps of both individual condensates and the cross term.
\begin{align} \notag
    I^{(1)} & = I_0^{(1)}(c_{1+},c_{1-}) +  I_\kappa^{(1)}
(c_{2+},c_{2-}) \\
    &  + I_\lambda^{(1)}(c_{1+},c_{1-},c_{2+},c_{2-}).
\end{align}
 The condensate overlap with its own trap reads simply $I_0^{(1)} = 4 \pi s_1 \vert H_1^{(1)} (kR) \vert^2$ where $s_{1,2} = |\mathbf{s}_{1,2}|$ are the pseudovector amplitudes of each condensate. We note that in standard tight-binding techniques the overlap integrals correspond to constant matrix terms $\langle \Psi | V |\Psi \rangle$ which lead to an eigenvalue problem whose solutions are the new energies and states of the coupled system. However, the ballistic nature of the polaritons is an interference problem, meaning that all integrals depend on the polariton outflow wavevector $k$ which is defined by the energy of the system. For this reason, the new energies of the system cannot be calculated from an eigenvalue problem unless one considers the interaction as a perturbation to the isolated condensate energy (see Eq.~\eqref{eq_resonance}). We will consider this perturbative tight-binding case in Sec.~\ref{sec:two}.
 
 The interference term reads,
\begin{align} \label{eq_I1}
    I_\lambda^{(1)} =& -\int_0^{2\pi} \left( c_{1+} e^{i\varphi} + c_{1-} e^{-i\varphi}\right) 
                      \left( c_{2+}^* e^{i\varphi^\prime} + c_{2-}^* e^{-i\varphi^\prime}\right)   \nonumber \\&
               \times  H_1^{(1)}(k R) H_1^{(1)}(k r^\prime)^* d \varphi + \textrm{c.c.}
\end{align}
The second integral due to the second condensate overlap with the first trap reads
\begin{equation}
    I_\kappa^{(1)} = \int_0^{2\pi} \left\vert c_{2+} e^{-i\varphi^\prime} + c_{2-} e^{i\varphi^\prime}\right\vert^2
                        \left\vert H_1^{(1)}(k r^\prime) \right\vert^2 d \varphi.
\end{equation}

With increasing distance between the traps $d/R$ the hierarchy of the terms $I_0^{(1)} \gg I_\lambda^{(1)} \gg I_\kappa^{(1)}$ quickly establishes due to exponential decay of the Hankel function. This allows us to neglect the last term $I_\kappa^{(1)}$ and treat the effect of interference $I_\lambda^{(1)}$ as perturbation to the first order with respect to decoupled condensates. The same argument obviously applies to the interference of the condensates at the second trap, with the corresponding integrals being $I_0^{(2)} = 4\pi s_2 \vert H_1^{(1)} (kR) \vert^2$ and,
\begin{align}
    I_\lambda^{(2)} =& \int_0^{2\pi} \left( c_{2+} e^{i\varphi} + c_{2-} e^{-i\varphi}\right)
                      \left( c_{1+}^* e^{-i\varphi^\prime} + c_{1-}^* e^{ i\varphi^\prime}\right)  \nonumber \\ & \times
                      H_1^{(1)}(k R) H_1^{(1)}(k r^\prime )^* d \varphi + \textrm{c.c.},
\end{align}
where $r^\prime = \sqrt{d^2 + R^2 + 2dR \cos(\varphi)}$ is slightly redefined as compared to the first trap case.

\subsection{Long distance limit}
\label{subsec:long}
Expanding Eqs.~\eqref{eq_rprime} and~\eqref{eq_phiprime} in $R/d \rightarrow 0$ we obtain in the integration range
\begin{equation}
\varphi^\prime \approx \sin(\varphi) {R \over d} \ll 1, \qquad
r^\prime \approx d \left(1 - {R \over d} \cos(\varphi) \right).
\end{equation}
In the main order in $R/d$ the interference integral \eqref{eq_I1} thus reads
\begin{align} \label{eq_I1_Rd} \nonumber
    I_\lambda^{(1)}= -\int_0^{2\pi} \left( c_{1+} e^{i\varphi} + c_{1-} e^{-i\varphi}\right) \times \\
                      \left[ c_{2+}^* \left(1 + i {R\over d} \sin(\varphi) \right) + c_{2-}^* \left(1-i{R\over d} \sin(\varphi)\right)\right] \times \\
         H_1^{(1)}(kR) H_1^{(1)}\left( kd - kR \cos(\varphi) \right)^* d \varphi + \textrm{c.c.} \nonumber
\end{align}
It is important to note here that $\vert k \vert \sim R^{-1}$ since the characteristic wavevector, corresponding to the highest state confined in a trap, is given by the inverse trap size 
We may thus asymptotically expand the Hankel function in Eq.~\eqref{eq_I1_Rd}:
\begin{align} \label{eq_H1}
    H_1^{(1)}\left( k[d - R \cos(\varphi)] \right) \approx {(1 + i) \over \sqrt{\pi k d}} \left[ 1 +{R \over 2d } \cos(\varphi) \right] \times \nonumber \\ \exp \left( i kd \right) \exp \left[ - i kR \cos(\varphi) \right]
\end{align}
Plugging \eqref{eq_H1} into \eqref{eq_I1_Rd} and employing the integral expression for Bessel function $2\pi i J_1(\rho) = \int_0^{2\pi} \cos(\varphi) \exp \left[ i\rho \cos(\varphi) \right]d \varphi$, we may compute the interference integral \eqref{eq_I1_Rd} to the leading order in $R/d$:
\begin{align}
    I_\lambda^{(1)} = -2 \pi i H_1^{(1)}(kR) \left( H_1^{(1)}(kd) J_1\left(kR \right) \right)^* \times \nonumber \\
    \left( c_{1+} + c_{1-} \right) \left( c_{2+} + c_{2-} \right)^*  + \textrm{c.c.}
\end{align}
The analogous interference integral for the first condensate overlap with the second trap reads
\begin{align}
    I_\lambda^{(2)} = -2 \pi i H_1^{(1)}(kR) \left( H_1^{(1)}(kd) J_1\left(kR \right) \right)^* \times \nonumber \\
    \left( c_{1+} + c_{1-} \right)^* \left( c_{2+} + c_{2-} \right)  + \textrm{c.c.},
\end{align}
therefore we have for the net interference contribution to the condensate overlap with the reservoir:
\begin{align} \label{eq_Sum}
    I_\text{int} = \ & I_\lambda^{(1)} + I_\lambda^{(2)} \nonumber \\ 
    = \ &  8 \pi \mathrm{Im} \left\lbrace H_1^{(1)}(kR) \left( H_1^{(1)}(kd) J_1\left(kR \right) \right)^* \right\rbrace \times \nonumber \\ 
   & \mathrm{Re}\left\lbrace \left( c_{1+} + c_{1-} \right) \left( c_{2+} + c_{2-} \right)^* \right\rbrace.
\end{align}

\subsection{General case}
\label{subsec:general}

\begin{figure}
    \centering
    \includegraphics[width=0.9\linewidth]{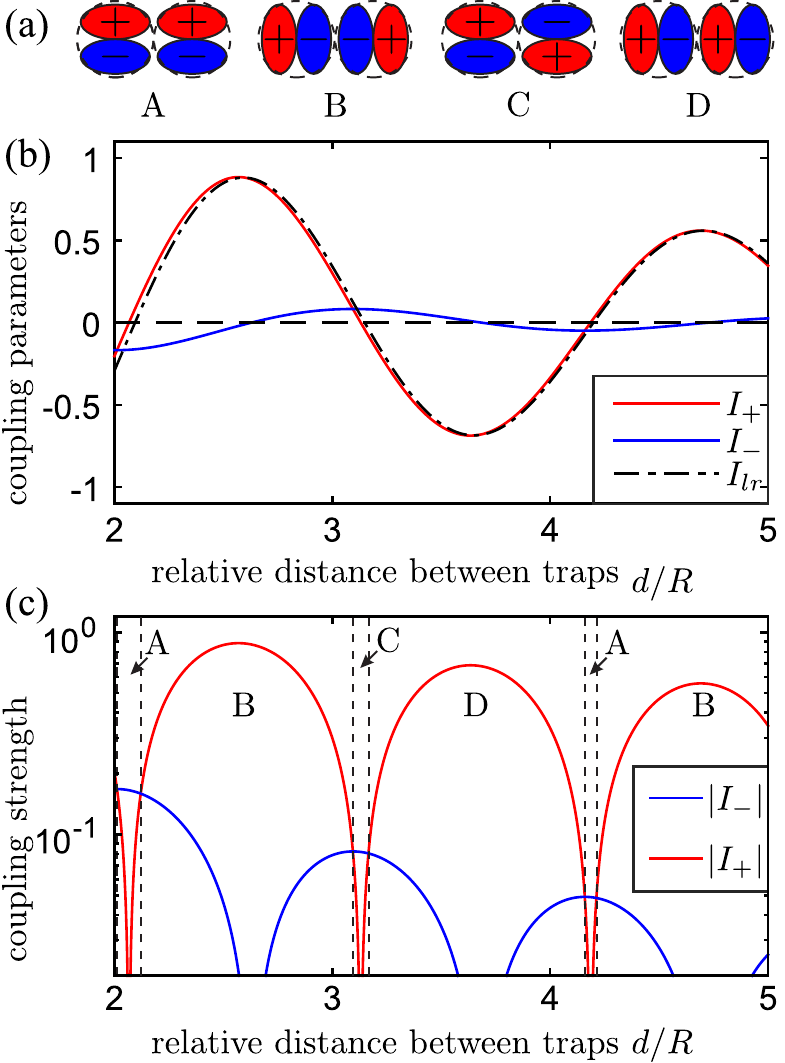}
    \caption{ (a) Illustration of the four possible coupling types. Regimes A and C share similarity with $\pi$-bonding, while B and D are similar to $\sigma$-bonds.
    (b) Coupling parameters for $|m|=1$ condensate interference for experimentally realistic parameter value $kR = 3+0.075i$. (c) The absolute values of coupling parameters. Switching of the coupling types occurs at the crossing points $\vert I_+ \vert = \vert I_- \vert$.}
    \label{fig3}
\end{figure}
\begin{figure}
    \centering
    \includegraphics[width=\linewidth]{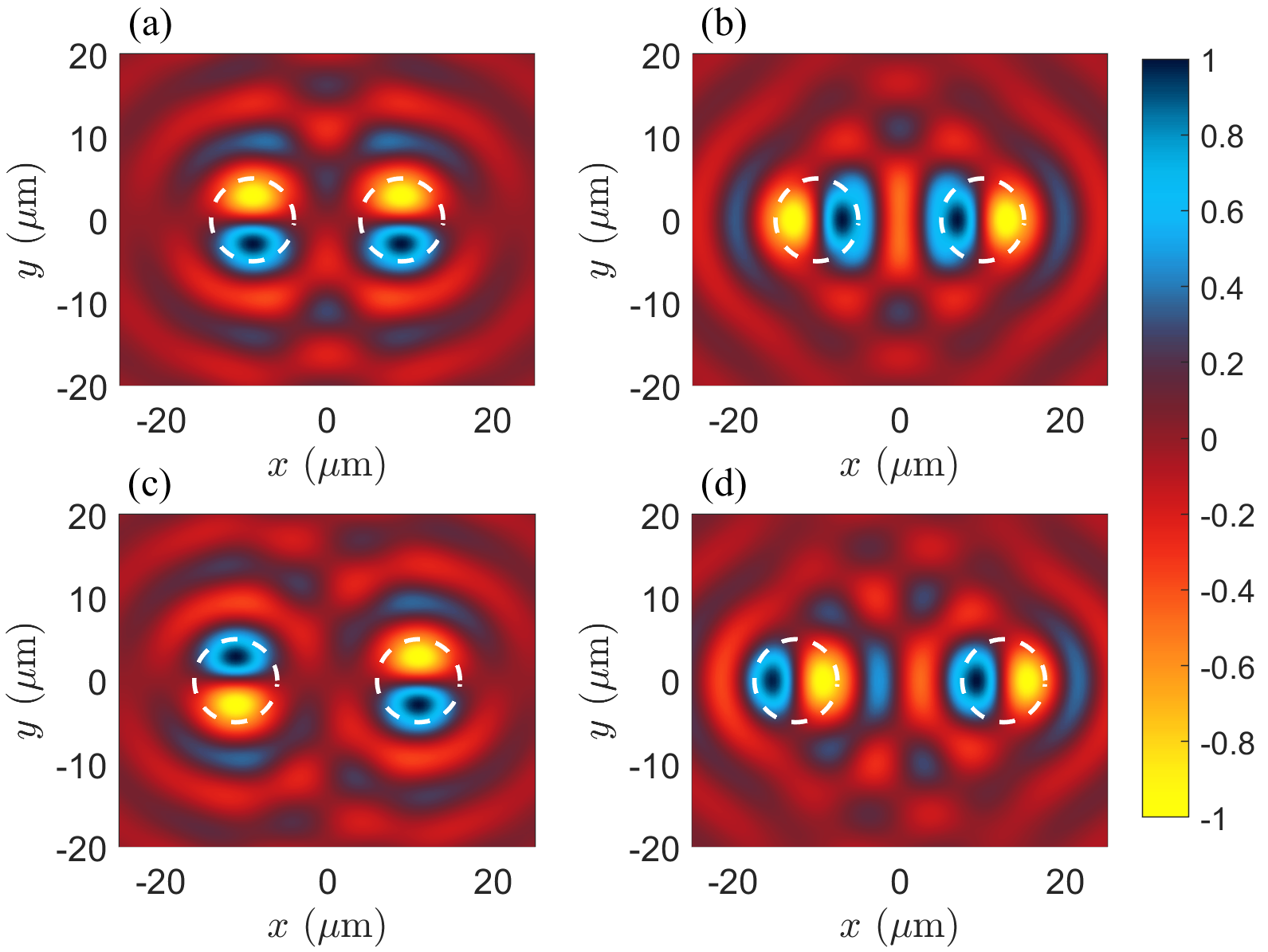}
    \caption{Steady state solutions $\text{Re}{[\Psi(\mathbf{r})]}$ from simulation of the dissipative Gross-Pitaevskii equation for two traps of $R = 5$ $\mu$m and separated by $d = 18, 20, 22, 25$ $\mu$m in (a)-(d) respectively. The states correspond to the configurations in the inset in Fig.~\ref{fig3}(a). White dashed circles denote the ridges of the traps. Panel (d) corresponds to the state shown in Fig.~\ref{fig2}(b).}
    \label{fig4}
\end{figure}

One may show for any distance $d$ between the traps that the overlap integral may be represented in the following form:
\begin{align} \label{eq_gen}
    I_\text{int} =& I_+ \mathrm{Re}\left\lbrace \left( c_{1+} + c_{1-} \right) \left( c_{2+} + c_{2-} \right)^* \right\rbrace  \nonumber \\ +& I_- \mathrm{Re}\left\lbrace \left( c_{1+} - c_{1-} \right) \left( c_{2+} - c_{2-} \right)^* \right\rbrace,
\end{align}
where $I_\pm = I_a \pm I_b$ and:
\begin{align}
    I_a =& \int_0^{2\pi} \left[
    e^{ i(\varphi_+^\prime - \varphi)} H_1^{(1)}(kR)^*H_1^{(1)}(kr_+^\prime)  \right. \nonumber \\
    & \left. -e^{ i(\varphi_-^\prime + \varphi)}H_1^{(1)}(kR)  H_1^{(1)}(kr_-^\prime)^* \right] d \varphi, \\
    I_b =& \int_0^{2\pi} \left[
    e^{ -i(\varphi + \varphi_+^\prime)}H_1^{(1)}(kR)^*H_1^{(1)}(kr_+^\prime)  \right. \nonumber \\ 
   & \left. -e^{ i(\varphi - \varphi_-^\prime)}H_1^{(1)}(kR)  H_1^{(1)}(kr_-^\prime)^* \right] d \varphi,
\end{align}
where $r_\pm^\prime= \sqrt{d^2 + R^2 \pm 2dR \cos(\varphi)}$ and $\varphi_\pm^\prime  = \sin^{-1}[(\sin(\varphi)R/r_\pm^\prime)]$. Physically, the integrals $I_\pm$ quantify the strength of interference coming from dipoles that are aligned longitudinally or transversely to the interference direction (the axis connecting the trap centers). Comparing Eqs.~\eqref{eq_Sum} and~\eqref{eq_gen} one may note that for long distances there exists an asymptotic relation.
\begin{equation}
    I_+(d) \approx I_{lr}(d) = 8 \pi \mathrm{Im} \left\lbrace H_1^{(1)}(kR) \left( H_1^{(1)}(kd) J_1\left(kR \right) \right)^* \right\rbrace,
\end{equation}
where $I_{lr}(d)$ is the long range part of the coupling parameter. The numerically computed values of the interaction parameters $I_\pm$ are shown in Fig.~\ref{fig3} for the experimentally realistic value of $kR = 3+0.075i$, corresponding to the following set of dimensional parameters: $m^*=5\times10^{-5}m_e$, $m_e$ being the free electron mass, $\Gamma = (10\, \text{ps})^{-1}$, $k = 1 + i0.025\,\mu$m$^{-1}$.
The asymptotic equivalence of computed the parameter $I_+(d)$ and $I_{lr}(d)$ is illustrated in Fig. \ref{fig3}a by the solid red and dot-dashed black lines respectively.
Moreover, in the logarithmic scale (Fig. \ref{fig3}b) it is clear that the envelope of $\vert I_+(d) \vert$ is exponentially dominant in the limit of long distances.
Nevertheless, the sign-changing and oscillating character of its dependence on the distance suggests that $\vert I_- \vert$, responsible for $\pi$ bonding type (regimes A and C), overcomes $\vert I_+ \vert$ in narrow regions of distances.

The net overlap integral may be represented in a more compact form $I = C^\dagger \hat{I} C$ in terms of the state vector $C = \left( c_{1+}, c_{1-}, c_{2+}, c_{2-} \right)^T$, where
\begin{equation} \label{eq_Itheta}
    \hat{I} = \left(\begin{matrix}
    0 & 0 & I_a & I_b e^{2i\theta} \\
    0 & 0 & I_b e^{-2i\theta} & I_a \\
    I_a & I_b e^{2i\theta} & 0 & 0 \\
    I_b e^{-2i\theta} & I_a & 0 & 0
    \end{matrix} \right).
\end{equation}
Here, we have accounted for an arbitrary angle $\theta$ between the axis, connecting the centers of the two traps, and the axis, from which the angles $\varphi$ and $\varphi^\prime$ are measured.
%

\section{The case of just two traps}
\label{sec:two}

The coupling derived in the previous section may be applied to various configurations of traps. The simplest case, illustrating the properties of the $p$-orbital dissipative coupling, is the case of only two traps, similar to the $s$-orbital coupling studied in~\cite{Harrison_PRB2020}. We compare our analytical results against simulations of the driven-dissipative Gross-Pistaevskii equation [Eqs.~\eqref{gpe} and~\eqref{res}].

In the particular case of only two coupled condensates we will write $c_{1\pm}=\sqrt{s_1}\exp\left[ i (\Phi_1 \mp \phi_1) \right]$, $c_{2\pm}=\sqrt{s_2}\exp\left[ i (\Phi_2 \mp \phi_2) \right]$, where $\Phi_{1,2}$ and $\phi_{1,2}$ define the phase of each condensate.
One then gets,
\begin{align} \notag
    \mathrm{Re}\left\lbrace \left( c_{1+} + c_{1-} \right)  \left( c_{2+} + c_{2-} \right)^* \right\rbrace & = \\ \label{eq_Rec}
    4 \sqrt{s_1 s_2}  \cos(\Delta \Phi) & \cos(\phi_1) \cos(\phi_2), \\ \notag
     \mathrm{Re}\left\lbrace \left( c_{1+} - c_{1-} \right) \left( c_{2+} - c_{2-} \right)^* \right\rbrace & = \\ \label{eq_Recm}
    4 \sqrt{s_1 s_2} \cos(\Delta \Phi) & \sin(\phi_1) \sin(\phi_2),
\end{align}
where $\Delta \Phi = \Phi_2 - \Phi_1$. Substituting Eqs.~\eqref{eq_Rec} and~\eqref{eq_Recm} into \eqref{eq_gen}, we obtain:
\begin{align} \notag
     I & = 4\sqrt{s_1 s_2} \cos(\Delta \Phi) \times \\  \label{eq_Ifin}
     & \left[ I_+ \cos(\phi_1) \cos(\phi_2) + I_- \sin(\phi_1) \sin(\phi_2) \right].
\end{align}
The relative ``external'' phase difference between the two condensates $\Delta \Phi$, as well as the ``internal'' phases $\phi_1$ and $\phi_2$ are chosen to maximize the value of the overlap integral \eqref{eq_Ifin}. 

We will consider the limiting case of the outflow wavevector $k= k_0$ being approximately unaffected by the coupling where $k_0$ is the wavevector corresponding to a single isolated condensate. The problem of identifying the solutions of the system is then analogous to a tight-binding theory where the new Hamiltonian describing the coupled system can be written,
\begin{equation} \label{eq.TB}
\hat{H} = \frac{V_0 \hat{I}(k_0)}{R}.
\end{equation}
We can directly infer from the relative strength and signs of the overlap integrals $I_\pm$, and the fact that $\text{Im}{(V_0)}>0$, which molecule configuration [see A-D in the inset of Fig. \ref{fig3}(a)] maximizes the gain of the system. If $\vert I_+ \vert > \vert I_- \vert$, then $\phi_1$ and $\phi_2$ take the values $\pi n$, corresponding to two dipole states aligned so that all four polariton density peaks lay on the line, connecting the centers of the traps. The relative phase $\Delta \Phi$ which leads to the strongest constructive interference and maximum gain is governed by the sign of $I_+$: negative sign corresponds to a antisymmetric wavefunction (B), while positive sign corresponds a symmetric wavefunction (D). If, on the contrary, $\vert I_+ \vert < \vert I_- \vert$, then $\phi_1$ and $\phi_2$ take the values $\pi (n+1/2)$, corresponding to both dipoles orientated transversely to the axis connecting the traps. Positive values of $I_-$ again correspond to a symmetric wavefunction (A), while negative values an antisymmetric wavefunction (C). The existence of all four configurations is confirmed through simulation of driven-dissipative Gross-Pitaevskii equation where in Fig. \ref{fig4} we show the resulting condensate steady state (fixed point solutions). We note that the $\pi$ solutions in Fig.~\ref{fig4}(a) and~\ref{fig4}(c) are unstable against fluctuations whereas the $\sigma$ solutions in Fig.~\ref{fig4}(b) and~\ref{fig4}(d) are stable. 
\begin{figure}
    \centering
    \includegraphics[width=\linewidth]{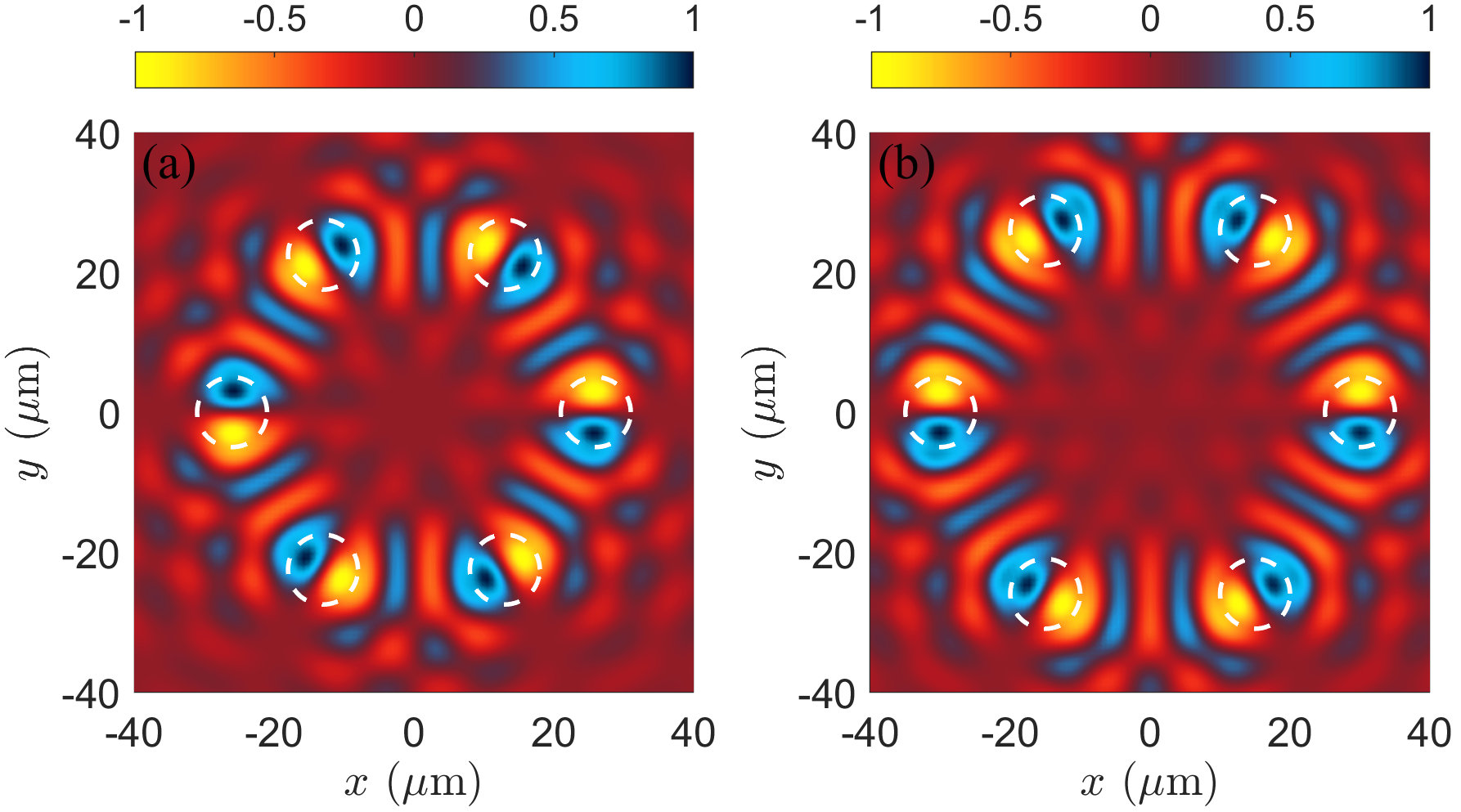}
    \caption{Two most stable configurations $\text{Re}{[\Psi(\mathbf{r})]}$ from simulation of the dissipative Gross-Pitaevskii equation for a hexagon of traps of $R = 5$ $\mu$m and circumradius $R_\text{hex} = 26 ,\ 30$ $\mu$m in (a,b) respectively. White dashed circles denote the ridges of the traps.}
    \label{fig5}
\end{figure}

We can notice that the condensate $\sigma$-bonds [configurations B and D in inset of Fig.~\ref{fig3}(a)] possess the lowest threshold over a much wider range compared to the $\pi$-bonds [configurations A and C in inset of Fig.~\ref{fig3}(a)]. In fact, this is analogous to molecular orbital theory where $\sigma$-bonds form the strongest covalent bonds, determined by the amount of overlap between the atomic orbitals. We additionally show results in Fig.~\ref{fig5} for a hexagon of coupled dipoles emulating a $\sigma$ bonded benzene-type molecule at two different molecule radii, $R_\text{hex} = 26, \ 30$ $\mu$m. The $\sigma$ bonded states are dominantly stable and we expect that such configurations will appear in experiment. To favour the $\pi$ bonding mechanism, one can introduce eccentricity $<1$ into the traps such that mode-splitting occurs and the condensate will start favoring $\pi$ bonded states. Indeed, such control over the trap geometry is easily possible in experiment and would allow for much more complex molecule simulation than shown in Fig.~\ref{fig4} and~\ref{fig5}.



\section{Conclusion}
We have proposed and investigated a system of exciton-polariton condensates for emulation of atomic orbitals, and taken steps towards realization of optical-based simulation of interacting condensates forming two-dimensional molecules. Currently, we have focused on the first excited $p$-orbital state in a cylindrically symmetric trap, already observed in a number of polariton experiments~\cite{Dreismann_PNAS2014, Gao_PRL2018, askitopoulos_all-optical_2018, Ma_NatCom2020, tpfer2020lotkavolterra}. From there, we studied the formation of in-phase and anti-phase $\pi$ and $\sigma$ type bonding configurations dictated by the system "optimal gain", or "lowest threshold", condition due to the dissipative nature of the polaritons. Interestingly, we find that the condensate $\sigma$ bonds are dominant, analogous to molecular bonding theory where they form the strongest covalent bonds and most stable molecular configuration. Our trap, which mimics the atomic potential, can be engineered through optical means, allowing for reprogrammable parameters of the potential such as its shape and size. The approach is straightforwardly extendable for the creation of lattices of artificial atoms with independent control over individual trap parameters. This allows for the implementation of a number of different geometries such as square, triagonal, honeycomb and more, which underscores the capabilities of the platform as a potential analogue simulator for conjugated systems of $p$-orbitals and possibly more complex molecules. The condensate occupation number scales with the excitation power which allows control over the influence of on-site particle interactions. The spin degree of freedom of the condensates can also be harnessed by controlling the excitation polarization, which can be used to explore the spinor states of the designed molecules~\cite{askitopoulos_all-optical_2018}. 

Beyond simulation of electron dynamics in molecules, our study also brings in the possibility to investigate networks of dissipative polariton condensates with directional (anisotropic) coupling mechanism. While $p$-orbital interactions of polariton condensates are already well established in high quality patterned microcavities~\cite{Galbiati_PRL2012, Mangussi_IOP2020}, networks of optical traps containing dissipatively coupled excited condensate orbitals have not been considered before. Of interest, by designing traps with open space between their centers like shown in Fig.~\ref{fig2}(b), as opposed to traps that are tightly packed, one can introduce the effect of time lag into the condensate coupling due to the finite travel time of particles. Such time delay coupling~\cite{Topfer_ComPhys2020}, available in our system, cannot be reproduced in patterned microcavities or cold atoms which rely on evanescent coupling (tunneling).

\section{Acknowledgements}
H.S. acknowledges the support of the UK’s Engineering and Physical Sciences Research Council (grant EP/M025330/1 on Hybrid Polaritonics) and hospitality provided by University of Iceland.
A.V.N. acknowledges support from European Union’s Horizon 2020 research and innovation programme under the Marie Skłodowska-Curie grant agreement No 846353.

\bibliography{main}

\begin{thebibliography}{44}%
\makeatletter
\providecommand \@ifxundefined [1]{%
 \@ifx{#1\undefined}
}%
\providecommand \@ifnum [1]{%
 \ifnum #1\expandafter \@firstoftwo
 \else \expandafter \@secondoftwo
 \fi
}%
\providecommand \@ifx [1]{%
 \ifx #1\expandafter \@firstoftwo
 \else \expandafter \@secondoftwo
 \fi
}%
\providecommand \natexlab [1]{#1}%
\providecommand \enquote  [1]{``#1''}%
\providecommand \bibnamefont  [1]{#1}%
\providecommand \bibfnamefont [1]{#1}%
\providecommand \citenamefont [1]{#1}%
\providecommand \href@noop [0]{\@secondoftwo}%
\providecommand \href [0]{\begingroup \@sanitize@url \@href}%
\providecommand \@href[1]{\@@startlink{#1}\@@href}%
\providecommand \@@href[1]{\endgroup#1\@@endlink}%
\providecommand \@sanitize@url [0]{\catcode `\\12\catcode `\$12\catcode
  `\&12\catcode `\#12\catcode `\^12\catcode `\_12\catcode `\%12\relax}%
\providecommand \@@startlink[1]{}%
\providecommand \@@endlink[0]{}%
\providecommand \url  [0]{\begingroup\@sanitize@url \@url }%
\providecommand \@url [1]{\endgroup\@href {#1}{\urlprefix }}%
\providecommand \urlprefix  [0]{URL }%
\providecommand \Eprint [0]{\href }%
\providecommand \doibase [0]{https://doi.org/}%
\providecommand \selectlanguage [0]{\@gobble}%
\providecommand \bibinfo  [0]{\@secondoftwo}%
\providecommand \bibfield  [0]{\@secondoftwo}%
\providecommand \translation [1]{[#1]}%
\providecommand \BibitemOpen [0]{}%
\providecommand \bibitemStop [0]{}%
\providecommand \bibitemNoStop [0]{.\EOS\space}%
\providecommand \EOS [0]{\spacefactor3000\relax}%
\providecommand \BibitemShut  [1]{\csname bibitem#1\endcsname}%
\let\auto@bib@innerbib\@empty
\bibitem [{\citenamefont {Georgescu}\ \emph {et~al.}(2014)\citenamefont
  {Georgescu}, \citenamefont {Ashhab},\ and\ \citenamefont
  {Nori}}]{Georgescu_RMP2014}%
  \BibitemOpen
  \bibfield  {author} {\bibinfo {author} {\bibfnamefont {I.~M.}\ \bibnamefont
  {Georgescu}}, \bibinfo {author} {\bibfnamefont {S.}~\bibnamefont {Ashhab}},\
  and\ \bibinfo {author} {\bibfnamefont {F.}~\bibnamefont {Nori}},\ }\bibfield
  {title} {\bibinfo {title} {Quantum simulation},\ }\href
  {https://doi.org/10.1103/RevModPhys.86.153} {\bibfield  {journal} {\bibinfo
  {journal} {Rev. Mod. Phys.}\ }\textbf {\bibinfo {volume} {86}},\ \bibinfo
  {pages} {153} (\bibinfo {year} {2014})}\BibitemShut {NoStop}%
\bibitem [{\citenamefont {Aspuru-Guzik}\ \emph {et~al.}(2005)\citenamefont
  {Aspuru-Guzik}, \citenamefont {Dutoi}, \citenamefont {Love},\ and\
  \citenamefont {Head-Gordon}}]{Aspuru_Science2005}%
  \BibitemOpen
  \bibfield  {author} {\bibinfo {author} {\bibfnamefont {A.}~\bibnamefont
  {Aspuru-Guzik}}, \bibinfo {author} {\bibfnamefont {A.~D.}\ \bibnamefont
  {Dutoi}}, \bibinfo {author} {\bibfnamefont {P.~J.}\ \bibnamefont {Love}},\
  and\ \bibinfo {author} {\bibfnamefont {M.}~\bibnamefont {Head-Gordon}},\
  }\bibfield  {title} {\bibinfo {title} {Simulated quantum computation of
  molecular energies},\ }\href {https://doi.org/10.1126/science.1113479}
  {\bibfield  {journal} {\bibinfo  {journal} {Science}\ }\textbf {\bibinfo
  {volume} {309}},\ \bibinfo {pages} {1704} (\bibinfo {year}
  {2005})}\BibitemShut {NoStop}%
\bibitem [{\citenamefont {O'Malley}\ \emph {et~al.}(2016)\citenamefont
  {O'Malley}, \citenamefont {Babbush}, \citenamefont {Kivlichan}, \citenamefont
  {Romero}, \citenamefont {McClean}, \citenamefont {Barends}, \citenamefont
  {Kelly}, \citenamefont {Roushan}, \citenamefont {Tranter}, \citenamefont
  {Ding}, \citenamefont {Campbell}, \citenamefont {Chen}, \citenamefont {Chen},
  \citenamefont {Chiaro}, \citenamefont {Dunsworth}, \citenamefont {Fowler},
  \citenamefont {Jeffrey}, \citenamefont {Lucero}, \citenamefont {Megrant},
  \citenamefont {Mutus}, \citenamefont {Neeley}, \citenamefont {Neill},
  \citenamefont {Quintana}, \citenamefont {Sank}, \citenamefont {Vainsencher},
  \citenamefont {Wenner}, \citenamefont {White}, \citenamefont {Coveney},
  \citenamefont {Love}, \citenamefont {Neven}, \citenamefont {Aspuru-Guzik},\
  and\ \citenamefont {Martinis}}]{Malley_PRX2016}%
  \BibitemOpen
  \bibfield  {author} {\bibinfo {author} {\bibfnamefont {P.~J.~J.}\
  \bibnamefont {O'Malley}}, \bibinfo {author} {\bibfnamefont {R.}~\bibnamefont
  {Babbush}}, \bibinfo {author} {\bibfnamefont {I.~D.}\ \bibnamefont
  {Kivlichan}}, \bibinfo {author} {\bibfnamefont {J.}~\bibnamefont {Romero}},
  \bibinfo {author} {\bibfnamefont {J.~R.}\ \bibnamefont {McClean}}, \bibinfo
  {author} {\bibfnamefont {R.}~\bibnamefont {Barends}}, \bibinfo {author}
  {\bibfnamefont {J.}~\bibnamefont {Kelly}}, \bibinfo {author} {\bibfnamefont
  {P.}~\bibnamefont {Roushan}}, \bibinfo {author} {\bibfnamefont
  {A.}~\bibnamefont {Tranter}}, \bibinfo {author} {\bibfnamefont
  {N.}~\bibnamefont {Ding}}, \bibinfo {author} {\bibfnamefont {B.}~\bibnamefont
  {Campbell}}, \bibinfo {author} {\bibfnamefont {Y.}~\bibnamefont {Chen}},
  \bibinfo {author} {\bibfnamefont {Z.}~\bibnamefont {Chen}}, \bibinfo {author}
  {\bibfnamefont {B.}~\bibnamefont {Chiaro}}, \bibinfo {author} {\bibfnamefont
  {A.}~\bibnamefont {Dunsworth}}, \bibinfo {author} {\bibfnamefont {A.~G.}\
  \bibnamefont {Fowler}}, \bibinfo {author} {\bibfnamefont {E.}~\bibnamefont
  {Jeffrey}}, \bibinfo {author} {\bibfnamefont {E.}~\bibnamefont {Lucero}},
  \bibinfo {author} {\bibfnamefont {A.}~\bibnamefont {Megrant}}, \bibinfo
  {author} {\bibfnamefont {J.~Y.}\ \bibnamefont {Mutus}}, \bibinfo {author}
  {\bibfnamefont {M.}~\bibnamefont {Neeley}}, \bibinfo {author} {\bibfnamefont
  {C.}~\bibnamefont {Neill}}, \bibinfo {author} {\bibfnamefont
  {C.}~\bibnamefont {Quintana}}, \bibinfo {author} {\bibfnamefont
  {D.}~\bibnamefont {Sank}}, \bibinfo {author} {\bibfnamefont {A.}~\bibnamefont
  {Vainsencher}}, \bibinfo {author} {\bibfnamefont {J.}~\bibnamefont {Wenner}},
  \bibinfo {author} {\bibfnamefont {T.~C.}\ \bibnamefont {White}}, \bibinfo
  {author} {\bibfnamefont {P.~V.}\ \bibnamefont {Coveney}}, \bibinfo {author}
  {\bibfnamefont {P.~J.}\ \bibnamefont {Love}}, \bibinfo {author}
  {\bibfnamefont {H.}~\bibnamefont {Neven}}, \bibinfo {author} {\bibfnamefont
  {A.}~\bibnamefont {Aspuru-Guzik}},\ and\ \bibinfo {author} {\bibfnamefont
  {J.~M.}\ \bibnamefont {Martinis}},\ }\bibfield  {title} {\bibinfo {title}
  {Scalable quantum simulation of molecular energies},\ }\href
  {https://doi.org/10.1103/PhysRevX.6.031007} {\bibfield  {journal} {\bibinfo
  {journal} {Phys. Rev. X}\ }\textbf {\bibinfo {volume} {6}},\ \bibinfo {pages}
  {031007} (\bibinfo {year} {2016})}\BibitemShut {NoStop}%
\bibitem [{\citenamefont {Kandala}\ \emph {et~al.}(2017)\citenamefont
  {Kandala}, \citenamefont {Mezzacapo}, \citenamefont {Temme}, \citenamefont
  {Takita}, \citenamefont {Brink}, \citenamefont {Chow},\ and\ \citenamefont
  {Gambetta}}]{Kandala_Nature2017}%
  \BibitemOpen
  \bibfield  {author} {\bibinfo {author} {\bibfnamefont {A.}~\bibnamefont
  {Kandala}}, \bibinfo {author} {\bibfnamefont {A.}~\bibnamefont {Mezzacapo}},
  \bibinfo {author} {\bibfnamefont {K.}~\bibnamefont {Temme}}, \bibinfo
  {author} {\bibfnamefont {M.}~\bibnamefont {Takita}}, \bibinfo {author}
  {\bibfnamefont {M.}~\bibnamefont {Brink}}, \bibinfo {author} {\bibfnamefont
  {J.~M.}\ \bibnamefont {Chow}},\ and\ \bibinfo {author} {\bibfnamefont
  {J.~M.}\ \bibnamefont {Gambetta}},\ }\bibfield  {title} {\bibinfo {title}
  {Hardware-efficient variational quantum eigensolver for small molecules and
  quantum magnets},\ }\href {https://doi.org/10.1038/nature23879} {\bibfield
  {journal} {\bibinfo  {journal} {Nature}\ }\textbf {\bibinfo {volume} {549}},\
  \bibinfo {pages} {242} (\bibinfo {year} {2017})}\BibitemShut {NoStop}%
\bibitem [{Aru(2020)}]{Arute_Science2020}%
  \BibitemOpen
  \bibfield  {title} {\bibinfo {title} {Hartree-fock on a superconducting qubit
  quantum computer},\ }\href {https://doi.org/10.1126/science.abb9811}
  {\bibfield  {journal} {\bibinfo  {journal} {Science}\ }\textbf {\bibinfo
  {volume} {369}},\ \bibinfo {pages} {1084} (\bibinfo {year}
  {2020})}\BibitemShut {NoStop}%
\bibitem [{\citenamefont {Lanyon}\ \emph {et~al.}(2010)\citenamefont {Lanyon},
  \citenamefont {Whitfield}, \citenamefont {Gillett}, \citenamefont {Goggin},
  \citenamefont {Almeida}, \citenamefont {Kassal}, \citenamefont {Biamonte},
  \citenamefont {Mohseni}, \citenamefont {Powell}, \citenamefont {Barbieri},
  \citenamefont {Aspuru-Guzik},\ and\ \citenamefont
  {White}}]{Lanyon_NatChem2010}%
  \BibitemOpen
  \bibfield  {author} {\bibinfo {author} {\bibfnamefont {B.~P.}\ \bibnamefont
  {Lanyon}}, \bibinfo {author} {\bibfnamefont {J.~D.}\ \bibnamefont
  {Whitfield}}, \bibinfo {author} {\bibfnamefont {G.~G.}\ \bibnamefont
  {Gillett}}, \bibinfo {author} {\bibfnamefont {M.~E.}\ \bibnamefont {Goggin}},
  \bibinfo {author} {\bibfnamefont {M.~P.}\ \bibnamefont {Almeida}}, \bibinfo
  {author} {\bibfnamefont {I.}~\bibnamefont {Kassal}}, \bibinfo {author}
  {\bibfnamefont {J.~D.}\ \bibnamefont {Biamonte}}, \bibinfo {author}
  {\bibfnamefont {M.}~\bibnamefont {Mohseni}}, \bibinfo {author} {\bibfnamefont
  {B.~J.}\ \bibnamefont {Powell}}, \bibinfo {author} {\bibfnamefont
  {M.}~\bibnamefont {Barbieri}}, \bibinfo {author} {\bibfnamefont
  {A.}~\bibnamefont {Aspuru-Guzik}},\ and\ \bibinfo {author} {\bibfnamefont
  {A.~G.}\ \bibnamefont {White}},\ }\bibfield  {title} {\bibinfo {title}
  {Towards quantum chemistry on a quantum computer},\ }\href
  {https://doi.org/10.1038/nchem.483} {\bibfield  {journal} {\bibinfo
  {journal} {Nature Chemistry}\ }\textbf {\bibinfo {volume} {2}},\ \bibinfo
  {pages} {106} (\bibinfo {year} {2010})}\BibitemShut {NoStop}%
\bibitem [{\citenamefont {Peruzzo}\ \emph {et~al.}(2014)\citenamefont
  {Peruzzo}, \citenamefont {McClean}, \citenamefont {Shadbolt}, \citenamefont
  {Yung}, \citenamefont {Zhou}, \citenamefont {Love}, \citenamefont
  {Aspuru-Guzik},\ and\ \citenamefont {O'Brien}}]{Peruzzo_NatComm2014}%
  \BibitemOpen
  \bibfield  {author} {\bibinfo {author} {\bibfnamefont {A.}~\bibnamefont
  {Peruzzo}}, \bibinfo {author} {\bibfnamefont {J.}~\bibnamefont {McClean}},
  \bibinfo {author} {\bibfnamefont {P.}~\bibnamefont {Shadbolt}}, \bibinfo
  {author} {\bibfnamefont {M.-H.}\ \bibnamefont {Yung}}, \bibinfo {author}
  {\bibfnamefont {X.-Q.}\ \bibnamefont {Zhou}}, \bibinfo {author}
  {\bibfnamefont {P.~J.}\ \bibnamefont {Love}}, \bibinfo {author}
  {\bibfnamefont {A.}~\bibnamefont {Aspuru-Guzik}},\ and\ \bibinfo {author}
  {\bibfnamefont {J.~L.}\ \bibnamefont {O'Brien}},\ }\bibfield  {title}
  {\bibinfo {title} {A variational eigenvalue solver on a photonic quantum
  processor},\ }\href {https://doi.org/10.1038/ncomms5213} {\bibfield
  {journal} {\bibinfo  {journal} {Nature Communications}\ }\textbf {\bibinfo
  {volume} {5}},\ \bibinfo {pages} {4213} (\bibinfo {year} {2014})}\BibitemShut
  {NoStop}%
\bibitem [{\citenamefont {L\"uhmann}\ \emph {et~al.}(2015)\citenamefont
  {L\"uhmann}, \citenamefont {Weitenberg},\ and\ \citenamefont
  {Sengstock}}]{Luhmann_PRX2015}%
  \BibitemOpen
  \bibfield  {author} {\bibinfo {author} {\bibfnamefont {D.-S.}\ \bibnamefont
  {L\"uhmann}}, \bibinfo {author} {\bibfnamefont {C.}~\bibnamefont
  {Weitenberg}},\ and\ \bibinfo {author} {\bibfnamefont {K.}~\bibnamefont
  {Sengstock}},\ }\bibfield  {title} {\bibinfo {title} {Emulating molecular
  orbitals and electronic dynamics with ultracold atoms},\ }\href
  {https://doi.org/10.1103/PhysRevX.5.031016} {\bibfield  {journal} {\bibinfo
  {journal} {Phys. Rev. X}\ }\textbf {\bibinfo {volume} {5}},\ \bibinfo {pages}
  {031016} (\bibinfo {year} {2015})}\BibitemShut {NoStop}%
\bibitem [{\citenamefont {Arg{\"u}ello-Luengo}\ \emph
  {et~al.}(2019)\citenamefont {Arg{\"u}ello-Luengo}, \citenamefont
  {Gonz{\'a}lez-Tudela}, \citenamefont {Shi}, \citenamefont {Zoller},\ and\
  \citenamefont {Cirac}}]{Arguello_Nature2019}%
  \BibitemOpen
  \bibfield  {author} {\bibinfo {author} {\bibfnamefont {J.}~\bibnamefont
  {Arg{\"u}ello-Luengo}}, \bibinfo {author} {\bibfnamefont {A.}~\bibnamefont
  {Gonz{\'a}lez-Tudela}}, \bibinfo {author} {\bibfnamefont {T.}~\bibnamefont
  {Shi}}, \bibinfo {author} {\bibfnamefont {P.}~\bibnamefont {Zoller}},\ and\
  \bibinfo {author} {\bibfnamefont {J.~I.}\ \bibnamefont {Cirac}},\ }\bibfield
  {title} {\bibinfo {title} {Analogue quantum chemistry simulation},\ }\href
  {https://doi.org/10.1038/s41586-019-1614-4} {\bibfield  {journal} {\bibinfo
  {journal} {Nature}\ }\textbf {\bibinfo {volume} {574}},\ \bibinfo {pages}
  {215} (\bibinfo {year} {2019})}\BibitemShut {NoStop}%
\bibitem [{\citenamefont {Kaitouni}\ \emph {et~al.}(2006)\citenamefont
  {Kaitouni}, \citenamefont {El~Da\"{\i}f}, \citenamefont {Baas}, \citenamefont
  {Richard}, \citenamefont {Paraiso}, \citenamefont {Lugan}, \citenamefont
  {Guillet}, \citenamefont {Morier-Genoud}, \citenamefont {Gani\`ere},
  \citenamefont {Staehli}, \citenamefont {Savona},\ and\ \citenamefont
  {Deveaud}}]{kaitouni_engineering_2006}%
  \BibitemOpen
  \bibfield  {author} {\bibinfo {author} {\bibfnamefont {R.~I.}\ \bibnamefont
  {Kaitouni}}, \bibinfo {author} {\bibfnamefont {O.}~\bibnamefont
  {El~Da\"{\i}f}}, \bibinfo {author} {\bibfnamefont {A.}~\bibnamefont {Baas}},
  \bibinfo {author} {\bibfnamefont {M.}~\bibnamefont {Richard}}, \bibinfo
  {author} {\bibfnamefont {T.}~\bibnamefont {Paraiso}}, \bibinfo {author}
  {\bibfnamefont {P.}~\bibnamefont {Lugan}}, \bibinfo {author} {\bibfnamefont
  {T.}~\bibnamefont {Guillet}}, \bibinfo {author} {\bibfnamefont
  {F.}~\bibnamefont {Morier-Genoud}}, \bibinfo {author} {\bibfnamefont {J.~D.}\
  \bibnamefont {Gani\`ere}}, \bibinfo {author} {\bibfnamefont {J.~L.}\
  \bibnamefont {Staehli}}, \bibinfo {author} {\bibfnamefont {V.}~\bibnamefont
  {Savona}},\ and\ \bibinfo {author} {\bibfnamefont {B.}~\bibnamefont
  {Deveaud}},\ }\bibfield  {title} {\bibinfo {title} {Engineering the spatial
  confinement of exciton polaritons in semiconductors},\ }\href
  {https://doi.org/10.1103/PhysRevB.74.155311} {\bibfield  {journal} {\bibinfo
  {journal} {Phys. Rev. B}\ }\textbf {\bibinfo {volume} {74}},\ \bibinfo
  {pages} {155311} (\bibinfo {year} {2006})}\BibitemShut {NoStop}%
\bibitem [{\citenamefont {Bajoni}\ \emph {et~al.}(2008)\citenamefont {Bajoni},
  \citenamefont {Senellart}, \citenamefont {Wertz}, \citenamefont {Sagnes},
  \citenamefont {Miard}, \citenamefont {Lema\^{\i}tre},\ and\ \citenamefont
  {Bloch}}]{Bajoni_PRL2008}%
  \BibitemOpen
  \bibfield  {author} {\bibinfo {author} {\bibfnamefont {D.}~\bibnamefont
  {Bajoni}}, \bibinfo {author} {\bibfnamefont {P.}~\bibnamefont {Senellart}},
  \bibinfo {author} {\bibfnamefont {E.}~\bibnamefont {Wertz}}, \bibinfo
  {author} {\bibfnamefont {I.}~\bibnamefont {Sagnes}}, \bibinfo {author}
  {\bibfnamefont {A.}~\bibnamefont {Miard}}, \bibinfo {author} {\bibfnamefont
  {A.}~\bibnamefont {Lema\^{\i}tre}},\ and\ \bibinfo {author} {\bibfnamefont
  {J.}~\bibnamefont {Bloch}},\ }\bibfield  {title} {\bibinfo {title} {Polariton
  laser using single micropillar
  $\mathrm{GaAs}\mathrm{\text{\ensuremath{-}}}\mathrm{GaAlAs}$ semiconductor
  cavities},\ }\href {https://doi.org/10.1103/PhysRevLett.100.047401}
  {\bibfield  {journal} {\bibinfo  {journal} {Phys. Rev. Lett.}\ }\textbf
  {\bibinfo {volume} {100}},\ \bibinfo {pages} {047401} (\bibinfo {year}
  {2008})}\BibitemShut {NoStop}%
\bibitem [{\citenamefont {Ferrier}\ \emph {et~al.}(2011)\citenamefont
  {Ferrier}, \citenamefont {Wertz}, \citenamefont {Johne}, \citenamefont
  {Solnyshkov}, \citenamefont {Senellart}, \citenamefont {Sagnes},
  \citenamefont {Lema\^{\i}tre}, \citenamefont {Malpuech},\ and\ \citenamefont
  {Bloch}}]{Ferrier_PRL2011}%
  \BibitemOpen
  \bibfield  {author} {\bibinfo {author} {\bibfnamefont {L.}~\bibnamefont
  {Ferrier}}, \bibinfo {author} {\bibfnamefont {E.}~\bibnamefont {Wertz}},
  \bibinfo {author} {\bibfnamefont {R.}~\bibnamefont {Johne}}, \bibinfo
  {author} {\bibfnamefont {D.~D.}\ \bibnamefont {Solnyshkov}}, \bibinfo
  {author} {\bibfnamefont {P.}~\bibnamefont {Senellart}}, \bibinfo {author}
  {\bibfnamefont {I.}~\bibnamefont {Sagnes}}, \bibinfo {author} {\bibfnamefont
  {A.}~\bibnamefont {Lema\^{\i}tre}}, \bibinfo {author} {\bibfnamefont
  {G.}~\bibnamefont {Malpuech}},\ and\ \bibinfo {author} {\bibfnamefont
  {J.}~\bibnamefont {Bloch}},\ }\bibfield  {title} {\bibinfo {title}
  {Interactions in confined polariton condensates},\ }\href
  {https://doi.org/10.1103/PhysRevLett.106.126401} {\bibfield  {journal}
  {\bibinfo  {journal} {Phys. Rev. Lett.}\ }\textbf {\bibinfo {volume} {106}},\
  \bibinfo {pages} {126401} (\bibinfo {year} {2011})}\BibitemShut {NoStop}%
\bibitem [{\citenamefont {Galbiati}\ \emph {et~al.}(2012)\citenamefont
  {Galbiati}, \citenamefont {Ferrier}, \citenamefont {Solnyshkov},
  \citenamefont {Tanese}, \citenamefont {Wertz}, \citenamefont {Amo},
  \citenamefont {Abbarchi}, \citenamefont {Senellart}, \citenamefont {Sagnes},
  \citenamefont {Lema\^{\i}tre}, \citenamefont {Galopin}, \citenamefont
  {Malpuech},\ and\ \citenamefont {Bloch}}]{Galbiati_PRL2012}%
  \BibitemOpen
  \bibfield  {author} {\bibinfo {author} {\bibfnamefont {M.}~\bibnamefont
  {Galbiati}}, \bibinfo {author} {\bibfnamefont {L.}~\bibnamefont {Ferrier}},
  \bibinfo {author} {\bibfnamefont {D.~D.}\ \bibnamefont {Solnyshkov}},
  \bibinfo {author} {\bibfnamefont {D.}~\bibnamefont {Tanese}}, \bibinfo
  {author} {\bibfnamefont {E.}~\bibnamefont {Wertz}}, \bibinfo {author}
  {\bibfnamefont {A.}~\bibnamefont {Amo}}, \bibinfo {author} {\bibfnamefont
  {M.}~\bibnamefont {Abbarchi}}, \bibinfo {author} {\bibfnamefont
  {P.}~\bibnamefont {Senellart}}, \bibinfo {author} {\bibfnamefont
  {I.}~\bibnamefont {Sagnes}}, \bibinfo {author} {\bibfnamefont
  {A.}~\bibnamefont {Lema\^{\i}tre}}, \bibinfo {author} {\bibfnamefont
  {E.}~\bibnamefont {Galopin}}, \bibinfo {author} {\bibfnamefont
  {G.}~\bibnamefont {Malpuech}},\ and\ \bibinfo {author} {\bibfnamefont
  {J.}~\bibnamefont {Bloch}},\ }\bibfield  {title} {\bibinfo {title} {Polariton
  condensation in photonic molecules},\ }\href
  {https://doi.org/10.1103/PhysRevLett.108.126403} {\bibfield  {journal}
  {\bibinfo  {journal} {Phys. Rev. Lett.}\ }\textbf {\bibinfo {volume} {108}},\
  \bibinfo {pages} {126403} (\bibinfo {year} {2012})}\BibitemShut {NoStop}%
\bibitem [{\citenamefont {Sala}\ \emph {et~al.}(2015)\citenamefont {Sala},
  \citenamefont {Solnyshkov}, \citenamefont {Carusotto}, \citenamefont
  {Jacqmin}, \citenamefont {Lema\^{\i}tre}, \citenamefont
  {Ter\ifmmode~\mbox{\c{c}}\else \c{c}\fi{}as}, \citenamefont {Nalitov},
  \citenamefont {Abbarchi}, \citenamefont {Galopin}, \citenamefont {Sagnes},
  \citenamefont {Bloch}, \citenamefont {Malpuech},\ and\ \citenamefont
  {Amo}}]{Sala_PRX2015}%
  \BibitemOpen
  \bibfield  {author} {\bibinfo {author} {\bibfnamefont {V.~G.}\ \bibnamefont
  {Sala}}, \bibinfo {author} {\bibfnamefont {D.~D.}\ \bibnamefont
  {Solnyshkov}}, \bibinfo {author} {\bibfnamefont {I.}~\bibnamefont
  {Carusotto}}, \bibinfo {author} {\bibfnamefont {T.}~\bibnamefont {Jacqmin}},
  \bibinfo {author} {\bibfnamefont {A.}~\bibnamefont {Lema\^{\i}tre}}, \bibinfo
  {author} {\bibfnamefont {H.}~\bibnamefont {Ter\ifmmode~\mbox{\c{c}}\else
  \c{c}\fi{}as}}, \bibinfo {author} {\bibfnamefont {A.}~\bibnamefont
  {Nalitov}}, \bibinfo {author} {\bibfnamefont {M.}~\bibnamefont {Abbarchi}},
  \bibinfo {author} {\bibfnamefont {E.}~\bibnamefont {Galopin}}, \bibinfo
  {author} {\bibfnamefont {I.}~\bibnamefont {Sagnes}}, \bibinfo {author}
  {\bibfnamefont {J.}~\bibnamefont {Bloch}}, \bibinfo {author} {\bibfnamefont
  {G.}~\bibnamefont {Malpuech}},\ and\ \bibinfo {author} {\bibfnamefont
  {A.}~\bibnamefont {Amo}},\ }\bibfield  {title} {\bibinfo {title} {Spin-orbit
  coupling for photons and polaritons in microstructures},\ }\href
  {https://doi.org/10.1103/PhysRevX.5.011034} {\bibfield  {journal} {\bibinfo
  {journal} {Phys. Rev. X}\ }\textbf {\bibinfo {volume} {5}},\ \bibinfo {pages}
  {011034} (\bibinfo {year} {2015})}\BibitemShut {NoStop}%
\bibitem [{\citenamefont {Mangussi}\ \emph {et~al.}(2020)\citenamefont
  {Mangussi}, \citenamefont {Mili{\'{c}}evi{\'{c}}}, \citenamefont {Sagnes},
  \citenamefont {Gratiet}, \citenamefont {Harouri}, \citenamefont
  {Lema{\^{\i}}tre}, \citenamefont {Bloch}, \citenamefont {Amo},\ and\
  \citenamefont {Usaj}}]{Mangussi_IOP2020}%
  \BibitemOpen
  \bibfield  {author} {\bibinfo {author} {\bibfnamefont {F.}~\bibnamefont
  {Mangussi}}, \bibinfo {author} {\bibfnamefont {M.}~\bibnamefont
  {Mili{\'{c}}evi{\'{c}}}}, \bibinfo {author} {\bibfnamefont {I.}~\bibnamefont
  {Sagnes}}, \bibinfo {author} {\bibfnamefont {L.~L.}\ \bibnamefont {Gratiet}},
  \bibinfo {author} {\bibfnamefont {A.}~\bibnamefont {Harouri}}, \bibinfo
  {author} {\bibfnamefont {A.}~\bibnamefont {Lema{\^{\i}}tre}}, \bibinfo
  {author} {\bibfnamefont {J.}~\bibnamefont {Bloch}}, \bibinfo {author}
  {\bibfnamefont {A.}~\bibnamefont {Amo}},\ and\ \bibinfo {author}
  {\bibfnamefont {G.}~\bibnamefont {Usaj}},\ }\bibfield  {title} {\bibinfo
  {title} {Multi-orbital tight binding model for cavity-polariton lattices},\
  }\href {https://doi.org/10.1088/1361-648x/ab8524} {\bibfield  {journal}
  {\bibinfo  {journal} {Journal of Physics: Condensed Matter}\ }\textbf
  {\bibinfo {volume} {32}},\ \bibinfo {pages} {315402} (\bibinfo {year}
  {2020})}\BibitemShut {NoStop}%
\bibitem [{\citenamefont {Askitopoulos}\ \emph {et~al.}(2013)\citenamefont
  {Askitopoulos}, \citenamefont {Ohadi}, \citenamefont {Kavokin}, \citenamefont
  {Hatzopoulos}, \citenamefont {Savvidis},\ and\ \citenamefont
  {Lagoudakis}}]{Askitopoulos_PRB2013}%
  \BibitemOpen
  \bibfield  {author} {\bibinfo {author} {\bibfnamefont {A.}~\bibnamefont
  {Askitopoulos}}, \bibinfo {author} {\bibfnamefont {H.}~\bibnamefont {Ohadi}},
  \bibinfo {author} {\bibfnamefont {A.~V.}\ \bibnamefont {Kavokin}}, \bibinfo
  {author} {\bibfnamefont {Z.}~\bibnamefont {Hatzopoulos}}, \bibinfo {author}
  {\bibfnamefont {P.~G.}\ \bibnamefont {Savvidis}},\ and\ \bibinfo {author}
  {\bibfnamefont {P.~G.}\ \bibnamefont {Lagoudakis}},\ }\bibfield  {title}
  {\bibinfo {title} {Polariton condensation in an optically induced
  two-dimensional potential},\ }\href
  {https://doi.org/10.1103/PhysRevB.88.041308} {\bibfield  {journal} {\bibinfo
  {journal} {Phys. Rev. B}\ }\textbf {\bibinfo {volume} {88}},\ \bibinfo
  {pages} {041308} (\bibinfo {year} {2013})}\BibitemShut {NoStop}%
\bibitem [{\citenamefont {Cristofolini}\ \emph {et~al.}(2013)\citenamefont
  {Cristofolini}, \citenamefont {Dreismann}, \citenamefont {Christmann},
  \citenamefont {Franchetti}, \citenamefont {Berloff}, \citenamefont {Tsotsis},
  \citenamefont {Hatzopoulos}, \citenamefont {Savvidis},\ and\ \citenamefont
  {Baumberg}}]{Cristofolini_PRL2013}%
  \BibitemOpen
  \bibfield  {author} {\bibinfo {author} {\bibfnamefont {P.}~\bibnamefont
  {Cristofolini}}, \bibinfo {author} {\bibfnamefont {A.}~\bibnamefont
  {Dreismann}}, \bibinfo {author} {\bibfnamefont {G.}~\bibnamefont
  {Christmann}}, \bibinfo {author} {\bibfnamefont {G.}~\bibnamefont
  {Franchetti}}, \bibinfo {author} {\bibfnamefont {N.~G.}\ \bibnamefont
  {Berloff}}, \bibinfo {author} {\bibfnamefont {P.}~\bibnamefont {Tsotsis}},
  \bibinfo {author} {\bibfnamefont {Z.}~\bibnamefont {Hatzopoulos}}, \bibinfo
  {author} {\bibfnamefont {P.~G.}\ \bibnamefont {Savvidis}},\ and\ \bibinfo
  {author} {\bibfnamefont {J.~J.}\ \bibnamefont {Baumberg}},\ }\bibfield
  {title} {\bibinfo {title} {Optical superfluid phase transitions and trapping
  of polariton condensates},\ }\href
  {https://doi.org/10.1103/PhysRevLett.110.186403} {\bibfield  {journal}
  {\bibinfo  {journal} {Phys. Rev. Lett.}\ }\textbf {\bibinfo {volume} {110}},\
  \bibinfo {pages} {186403} (\bibinfo {year} {2013})}\BibitemShut {NoStop}%
\bibitem [{\citenamefont {Ohadi}\ \emph {et~al.}(2017)\citenamefont {Ohadi},
  \citenamefont {Ramsay}, \citenamefont {Sigurdsson}, \citenamefont {del
  Valle-Inclan~Redondo}, \citenamefont {Tsintzos}, \citenamefont {Hatzopoulos},
  \citenamefont {Liew}, \citenamefont {Shelykh}, \citenamefont {Rubo},
  \citenamefont {Savvidis},\ and\ \citenamefont {Baumberg}}]{Ohadi_PRL2017}%
  \BibitemOpen
  \bibfield  {author} {\bibinfo {author} {\bibfnamefont {H.}~\bibnamefont
  {Ohadi}}, \bibinfo {author} {\bibfnamefont {A.~J.}\ \bibnamefont {Ramsay}},
  \bibinfo {author} {\bibfnamefont {H.}~\bibnamefont {Sigurdsson}}, \bibinfo
  {author} {\bibfnamefont {Y.}~\bibnamefont {del Valle-Inclan~Redondo}},
  \bibinfo {author} {\bibfnamefont {S.~I.}\ \bibnamefont {Tsintzos}}, \bibinfo
  {author} {\bibfnamefont {Z.}~\bibnamefont {Hatzopoulos}}, \bibinfo {author}
  {\bibfnamefont {T.~C.~H.}\ \bibnamefont {Liew}}, \bibinfo {author}
  {\bibfnamefont {I.~A.}\ \bibnamefont {Shelykh}}, \bibinfo {author}
  {\bibfnamefont {Y.~G.}\ \bibnamefont {Rubo}}, \bibinfo {author}
  {\bibfnamefont {P.~G.}\ \bibnamefont {Savvidis}},\ and\ \bibinfo {author}
  {\bibfnamefont {J.~J.}\ \bibnamefont {Baumberg}},\ }\bibfield  {title}
  {\bibinfo {title} {Spin order and phase transitions in chains of polariton
  condensates},\ }\href {https://doi.org/10.1103/PhysRevLett.119.067401}
  {\bibfield  {journal} {\bibinfo  {journal} {Phys. Rev. Lett.}\ }\textbf
  {\bibinfo {volume} {119}},\ \bibinfo {pages} {067401} (\bibinfo {year}
  {2017})}\BibitemShut {NoStop}%
\bibitem [{\citenamefont {Gao}\ \emph {et~al.}(2015)\citenamefont {Gao},
  \citenamefont {Estrecho}, \citenamefont {Bliokh}, \citenamefont {Liew},
  \citenamefont {Fraser}, \citenamefont {Brodbeck}, \citenamefont {Kamp},
  \citenamefont {Schneider}, \citenamefont {H{\"o}fling}, \citenamefont
  {Yamamoto}, \citenamefont {Nori}, \citenamefont {Kivshar}, \citenamefont
  {Truscott}, \citenamefont {Dall},\ and\ \citenamefont
  {Ostrovskaya}}]{Gao_Nature2015}%
  \BibitemOpen
  \bibfield  {author} {\bibinfo {author} {\bibfnamefont {T.}~\bibnamefont
  {Gao}}, \bibinfo {author} {\bibfnamefont {E.}~\bibnamefont {Estrecho}},
  \bibinfo {author} {\bibfnamefont {K.~Y.}\ \bibnamefont {Bliokh}}, \bibinfo
  {author} {\bibfnamefont {T.~C.~H.}\ \bibnamefont {Liew}}, \bibinfo {author}
  {\bibfnamefont {M.~D.}\ \bibnamefont {Fraser}}, \bibinfo {author}
  {\bibfnamefont {S.}~\bibnamefont {Brodbeck}}, \bibinfo {author}
  {\bibfnamefont {M.}~\bibnamefont {Kamp}}, \bibinfo {author} {\bibfnamefont
  {C.}~\bibnamefont {Schneider}}, \bibinfo {author} {\bibfnamefont
  {S.}~\bibnamefont {H{\"o}fling}}, \bibinfo {author} {\bibfnamefont
  {Y.}~\bibnamefont {Yamamoto}}, \bibinfo {author} {\bibfnamefont
  {F.}~\bibnamefont {Nori}}, \bibinfo {author} {\bibfnamefont {Y.~S.}\
  \bibnamefont {Kivshar}}, \bibinfo {author} {\bibfnamefont {A.~G.}\
  \bibnamefont {Truscott}}, \bibinfo {author} {\bibfnamefont {R.~G.}\
  \bibnamefont {Dall}},\ and\ \bibinfo {author} {\bibfnamefont {E.~A.}\
  \bibnamefont {Ostrovskaya}},\ }\bibfield  {title} {\bibinfo {title}
  {Observation of non-hermitian degeneracies in a chaotic exciton-polariton
  billiard},\ }\href {https://doi.org/10.1038/nature15522} {\bibfield
  {journal} {\bibinfo  {journal} {Nature}\ }\textbf {\bibinfo {volume} {526}},\
  \bibinfo {pages} {554} (\bibinfo {year} {2015})}\BibitemShut {NoStop}%
\bibitem [{\citenamefont {Gao}\ \emph {et~al.}(2018)\citenamefont {Gao},
  \citenamefont {Li}, \citenamefont {Estrecho}, \citenamefont {Liew},
  \citenamefont {Comber-Todd}, \citenamefont {Nalitov}, \citenamefont {Steger},
  \citenamefont {West}, \citenamefont {Pfeiffer}, \citenamefont {Snoke},
  \citenamefont {Kavokin}, \citenamefont {Truscott},\ and\ \citenamefont
  {Ostrovskaya}}]{Gao_PRL2018}%
  \BibitemOpen
  \bibfield  {author} {\bibinfo {author} {\bibfnamefont {T.}~\bibnamefont
  {Gao}}, \bibinfo {author} {\bibfnamefont {G.}~\bibnamefont {Li}}, \bibinfo
  {author} {\bibfnamefont {E.}~\bibnamefont {Estrecho}}, \bibinfo {author}
  {\bibfnamefont {T.~C.~H.}\ \bibnamefont {Liew}}, \bibinfo {author}
  {\bibfnamefont {D.}~\bibnamefont {Comber-Todd}}, \bibinfo {author}
  {\bibfnamefont {A.}~\bibnamefont {Nalitov}}, \bibinfo {author} {\bibfnamefont
  {M.}~\bibnamefont {Steger}}, \bibinfo {author} {\bibfnamefont
  {K.}~\bibnamefont {West}}, \bibinfo {author} {\bibfnamefont {L.}~\bibnamefont
  {Pfeiffer}}, \bibinfo {author} {\bibfnamefont {D.~W.}\ \bibnamefont {Snoke}},
  \bibinfo {author} {\bibfnamefont {A.~V.}\ \bibnamefont {Kavokin}}, \bibinfo
  {author} {\bibfnamefont {A.~G.}\ \bibnamefont {Truscott}},\ and\ \bibinfo
  {author} {\bibfnamefont {E.~A.}\ \bibnamefont {Ostrovskaya}},\ }\bibfield
  {title} {\bibinfo {title} {Chiral modes at exceptional points in
  exciton-polariton quantum fluids},\ }\href
  {https://doi.org/10.1103/PhysRevLett.120.065301} {\bibfield  {journal}
  {\bibinfo  {journal} {Phys. Rev. Lett.}\ }\textbf {\bibinfo {volume} {120}},\
  \bibinfo {pages} {065301} (\bibinfo {year} {2018})}\BibitemShut {NoStop}%
\bibitem [{\citenamefont {Ma}\ \emph {et~al.}(2020)\citenamefont {Ma},
  \citenamefont {Berger}, \citenamefont {A{\ss}mann}, \citenamefont {Driben},
  \citenamefont {Meier}, \citenamefont {Schneider}, \citenamefont
  {H{\"o}fling},\ and\ \citenamefont {Schumacher}}]{Ma_NatCom2020}%
  \BibitemOpen
  \bibfield  {author} {\bibinfo {author} {\bibfnamefont {X.}~\bibnamefont
  {Ma}}, \bibinfo {author} {\bibfnamefont {B.}~\bibnamefont {Berger}}, \bibinfo
  {author} {\bibfnamefont {M.}~\bibnamefont {A{\ss}mann}}, \bibinfo {author}
  {\bibfnamefont {R.}~\bibnamefont {Driben}}, \bibinfo {author} {\bibfnamefont
  {T.}~\bibnamefont {Meier}}, \bibinfo {author} {\bibfnamefont
  {C.}~\bibnamefont {Schneider}}, \bibinfo {author} {\bibfnamefont
  {S.}~\bibnamefont {H{\"o}fling}},\ and\ \bibinfo {author} {\bibfnamefont
  {S.}~\bibnamefont {Schumacher}},\ }\bibfield  {title} {\bibinfo {title}
  {Realization of all-optical vortex switching in exciton-polariton
  condensates},\ }\href {https://doi.org/10.1038/s41467-020-14702-5} {\bibfield
   {journal} {\bibinfo  {journal} {Nature Communications}\ }\textbf {\bibinfo
  {volume} {11}},\ \bibinfo {pages} {897} (\bibinfo {year} {2020})}\BibitemShut
  {NoStop}%
\bibitem [{\citenamefont {Pickup}\ \emph {et~al.}(2018)\citenamefont {Pickup},
  \citenamefont {Kalinin}, \citenamefont {Askitopoulos}, \citenamefont
  {Hatzopoulos}, \citenamefont {Savvidis}, \citenamefont {Berloff},\ and\
  \citenamefont {Lagoudakis}}]{Pickup_PRL2018}%
  \BibitemOpen
  \bibfield  {author} {\bibinfo {author} {\bibfnamefont {L.}~\bibnamefont
  {Pickup}}, \bibinfo {author} {\bibfnamefont {K.}~\bibnamefont {Kalinin}},
  \bibinfo {author} {\bibfnamefont {A.}~\bibnamefont {Askitopoulos}}, \bibinfo
  {author} {\bibfnamefont {Z.}~\bibnamefont {Hatzopoulos}}, \bibinfo {author}
  {\bibfnamefont {P.~G.}\ \bibnamefont {Savvidis}}, \bibinfo {author}
  {\bibfnamefont {N.~G.}\ \bibnamefont {Berloff}},\ and\ \bibinfo {author}
  {\bibfnamefont {P.~G.}\ \bibnamefont {Lagoudakis}},\ }\bibfield  {title}
  {\bibinfo {title} {Optical bistability under nonresonant excitation in spinor
  polariton condensates},\ }\href
  {https://doi.org/10.1103/PhysRevLett.120.225301} {\bibfield  {journal}
  {\bibinfo  {journal} {Phys. Rev. Lett.}\ }\textbf {\bibinfo {volume} {120}},\
  \bibinfo {pages} {225301} (\bibinfo {year} {2018})}\BibitemShut {NoStop}%
\bibitem [{\citenamefont {Töpfer}\ \emph
  {et~al.}(2020{\natexlab{a}})\citenamefont {Töpfer}, \citenamefont
  {Sigurdsson}, \citenamefont {Alyatkin},\ and\ \citenamefont
  {Lagoudakis}}]{tpfer2020lotkavolterra}%
  \BibitemOpen
  \bibfield  {author} {\bibinfo {author} {\bibfnamefont {J.~D.}\ \bibnamefont
  {Töpfer}}, \bibinfo {author} {\bibfnamefont {H.}~\bibnamefont {Sigurdsson}},
  \bibinfo {author} {\bibfnamefont {S.}~\bibnamefont {Alyatkin}},\ and\
  \bibinfo {author} {\bibfnamefont {P.~G.}\ \bibnamefont {Lagoudakis}},\
  }\href@noop {} {\bibinfo {title} {Lotka-volterra population dynamics in
  coherent and tunable oscillators of trapped polariton condensates}} (\bibinfo
  {year} {2020}{\natexlab{a}}),\ \Eprint {https://arxiv.org/abs/2009.05637}
  {arXiv:2009.05637 [cond-mat.mes-hall]} \BibitemShut {NoStop}%
\bibitem [{\citenamefont {Töpfer}\ \emph
  {et~al.}(2020{\natexlab{b}})\citenamefont {Töpfer}, \citenamefont
  {Chatzopoulos}, \citenamefont {Sigurdsson}, \citenamefont {Cookson},
  \citenamefont {Rubo},\ and\ \citenamefont
  {Lagoudakis}}]{topfer_engineering_2020}%
  \BibitemOpen
  \bibfield  {author} {\bibinfo {author} {\bibfnamefont {J.~D.}\ \bibnamefont
  {Töpfer}}, \bibinfo {author} {\bibfnamefont {I.}~\bibnamefont
  {Chatzopoulos}}, \bibinfo {author} {\bibfnamefont {H.}~\bibnamefont
  {Sigurdsson}}, \bibinfo {author} {\bibfnamefont {T.}~\bibnamefont {Cookson}},
  \bibinfo {author} {\bibfnamefont {Y.~G.}\ \bibnamefont {Rubo}},\ and\
  \bibinfo {author} {\bibfnamefont {P.~G.}\ \bibnamefont {Lagoudakis}},\
  }\href@noop {} {\bibinfo {title} {Engineering spatial coherence in lattices
  of polariton condensates}} (\bibinfo {year} {2020}{\natexlab{b}}),\ \Eprint
  {https://arxiv.org/abs/2007.06690} {arXiv:2007.06690 [cond-mat.mes-hall]}
  \BibitemShut {NoStop}%
\bibitem [{\citenamefont {Maragkou}\ \emph {et~al.}(2010)\citenamefont
  {Maragkou}, \citenamefont {Grundy}, \citenamefont {Wertz}, \citenamefont
  {Lema\^{\i}tre}, \citenamefont {Sagnes}, \citenamefont {Senellart},
  \citenamefont {Bloch},\ and\ \citenamefont {Lagoudakis}}]{Maragkou_PRB2010}%
  \BibitemOpen
  \bibfield  {author} {\bibinfo {author} {\bibfnamefont {M.}~\bibnamefont
  {Maragkou}}, \bibinfo {author} {\bibfnamefont {A.~J.~D.}\ \bibnamefont
  {Grundy}}, \bibinfo {author} {\bibfnamefont {E.}~\bibnamefont {Wertz}},
  \bibinfo {author} {\bibfnamefont {A.}~\bibnamefont {Lema\^{\i}tre}}, \bibinfo
  {author} {\bibfnamefont {I.}~\bibnamefont {Sagnes}}, \bibinfo {author}
  {\bibfnamefont {P.}~\bibnamefont {Senellart}}, \bibinfo {author}
  {\bibfnamefont {J.}~\bibnamefont {Bloch}},\ and\ \bibinfo {author}
  {\bibfnamefont {P.~G.}\ \bibnamefont {Lagoudakis}},\ }\bibfield  {title}
  {\bibinfo {title} {Spontaneous nonground state polariton condensation in
  pillar microcavities},\ }\href {https://doi.org/10.1103/PhysRevB.81.081307}
  {\bibfield  {journal} {\bibinfo  {journal} {Phys. Rev. B}\ }\textbf {\bibinfo
  {volume} {81}},\ \bibinfo {pages} {081307} (\bibinfo {year}
  {2010})}\BibitemShut {NoStop}%
\bibitem [{\citenamefont {Manni}\ \emph {et~al.}(2011)\citenamefont {Manni},
  \citenamefont {Lagoudakis}, \citenamefont {Liew}, \citenamefont {Andr\'e},\
  and\ \citenamefont {Deveaud-Pl\'edran}}]{Manni_PRL2011}%
  \BibitemOpen
  \bibfield  {author} {\bibinfo {author} {\bibfnamefont {F.}~\bibnamefont
  {Manni}}, \bibinfo {author} {\bibfnamefont {K.~G.}\ \bibnamefont
  {Lagoudakis}}, \bibinfo {author} {\bibfnamefont {T.~C.~H.}\ \bibnamefont
  {Liew}}, \bibinfo {author} {\bibfnamefont {R.}~\bibnamefont {Andr\'e}},\ and\
  \bibinfo {author} {\bibfnamefont {B.}~\bibnamefont {Deveaud-Pl\'edran}},\
  }\bibfield  {title} {\bibinfo {title} {Spontaneous pattern formation in a
  polariton condensate},\ }\href
  {https://doi.org/10.1103/PhysRevLett.107.106401} {\bibfield  {journal}
  {\bibinfo  {journal} {Phys. Rev. Lett.}\ }\textbf {\bibinfo {volume} {107}},\
  \bibinfo {pages} {106401} (\bibinfo {year} {2011})}\BibitemShut {NoStop}%
\bibitem [{\citenamefont {Tosi}\ \emph {et~al.}(2012)\citenamefont {Tosi},
  \citenamefont {Christmann}, \citenamefont {Berloff}, \citenamefont {Tsotsis},
  \citenamefont {Gao}, \citenamefont {Hatzopoulos}, \citenamefont {Savvidis},\
  and\ \citenamefont {Baumberg}}]{Tosi_NatPhys2012}%
  \BibitemOpen
  \bibfield  {author} {\bibinfo {author} {\bibfnamefont {G.}~\bibnamefont
  {Tosi}}, \bibinfo {author} {\bibfnamefont {G.}~\bibnamefont {Christmann}},
  \bibinfo {author} {\bibfnamefont {N.~G.}\ \bibnamefont {Berloff}}, \bibinfo
  {author} {\bibfnamefont {P.}~\bibnamefont {Tsotsis}}, \bibinfo {author}
  {\bibfnamefont {T.}~\bibnamefont {Gao}}, \bibinfo {author} {\bibfnamefont
  {Z.}~\bibnamefont {Hatzopoulos}}, \bibinfo {author} {\bibfnamefont {P.~G.}\
  \bibnamefont {Savvidis}},\ and\ \bibinfo {author} {\bibfnamefont {J.~J.}\
  \bibnamefont {Baumberg}},\ }\bibfield  {title} {\bibinfo {title} {Sculpting
  oscillators with light within a nonlinear quantum fluid},\ }\href
  {https://doi.org/10.1038/nphys2182} {\bibfield  {journal} {\bibinfo
  {journal} {Nature Physics}\ }\textbf {\bibinfo {volume} {8}},\ \bibinfo
  {pages} {190} (\bibinfo {year} {2012})}\BibitemShut {NoStop}%
\bibitem [{\citenamefont {Dreismann}\ \emph {et~al.}(2014)\citenamefont
  {Dreismann}, \citenamefont {Cristofolini}, \citenamefont {Balili},
  \citenamefont {Christmann}, \citenamefont {Pinsker}, \citenamefont {Berloff},
  \citenamefont {Hatzopoulos}, \citenamefont {Savvidis},\ and\ \citenamefont
  {Baumberg}}]{Dreismann_PNAS2014}%
  \BibitemOpen
  \bibfield  {author} {\bibinfo {author} {\bibfnamefont {A.}~\bibnamefont
  {Dreismann}}, \bibinfo {author} {\bibfnamefont {P.}~\bibnamefont
  {Cristofolini}}, \bibinfo {author} {\bibfnamefont {R.}~\bibnamefont
  {Balili}}, \bibinfo {author} {\bibfnamefont {G.}~\bibnamefont {Christmann}},
  \bibinfo {author} {\bibfnamefont {F.}~\bibnamefont {Pinsker}}, \bibinfo
  {author} {\bibfnamefont {N.~G.}\ \bibnamefont {Berloff}}, \bibinfo {author}
  {\bibfnamefont {Z.}~\bibnamefont {Hatzopoulos}}, \bibinfo {author}
  {\bibfnamefont {P.~G.}\ \bibnamefont {Savvidis}},\ and\ \bibinfo {author}
  {\bibfnamefont {J.~J.}\ \bibnamefont {Baumberg}},\ }\bibfield  {title}
  {\bibinfo {title} {Coupled counterrotating polariton condensates in optically
  defined annular potentials},\ }\href
  {https://doi.org/10.1073/pnas.1401988111} {\bibfield  {journal} {\bibinfo
  {journal} {Proceedings of the National Academy of Sciences}\ }\textbf
  {\bibinfo {volume} {111}},\ \bibinfo {pages} {8770} (\bibinfo {year}
  {2014})}\BibitemShut {NoStop}%
\bibitem [{\citenamefont {Askitopoulos}\ \emph {et~al.}(2015)\citenamefont
  {Askitopoulos}, \citenamefont {Liew}, \citenamefont {Ohadi}, \citenamefont
  {Hatzopoulos}, \citenamefont {Savvidis},\ and\ \citenamefont
  {Lagoudakis}}]{Askitopoulos_PRB2015}%
  \BibitemOpen
  \bibfield  {author} {\bibinfo {author} {\bibfnamefont {A.}~\bibnamefont
  {Askitopoulos}}, \bibinfo {author} {\bibfnamefont {T.~C.~H.}\ \bibnamefont
  {Liew}}, \bibinfo {author} {\bibfnamefont {H.}~\bibnamefont {Ohadi}},
  \bibinfo {author} {\bibfnamefont {Z.}~\bibnamefont {Hatzopoulos}}, \bibinfo
  {author} {\bibfnamefont {P.~G.}\ \bibnamefont {Savvidis}},\ and\ \bibinfo
  {author} {\bibfnamefont {P.~G.}\ \bibnamefont {Lagoudakis}},\ }\bibfield
  {title} {\bibinfo {title} {Robust platform for engineering pure-quantum-state
  transitions in polariton condensates},\ }\href
  {https://doi.org/10.1103/PhysRevB.92.035305} {\bibfield  {journal} {\bibinfo
  {journal} {Phys. Rev. B}\ }\textbf {\bibinfo {volume} {92}},\ \bibinfo
  {pages} {035305} (\bibinfo {year} {2015})}\BibitemShut {NoStop}%
\bibitem [{\citenamefont {Sun}\ \emph {et~al.}(2018)\citenamefont {Sun},
  \citenamefont {Yoon}, \citenamefont {Khan}, \citenamefont {Ge}, \citenamefont
  {Steger}, \citenamefont {Pfeiffer}, \citenamefont {West}, \citenamefont
  {T\"ureci}, \citenamefont {Snoke},\ and\ \citenamefont
  {Nelson}}]{Sun_PRB2018}%
  \BibitemOpen
  \bibfield  {author} {\bibinfo {author} {\bibfnamefont {Y.}~\bibnamefont
  {Sun}}, \bibinfo {author} {\bibfnamefont {Y.}~\bibnamefont {Yoon}}, \bibinfo
  {author} {\bibfnamefont {S.}~\bibnamefont {Khan}}, \bibinfo {author}
  {\bibfnamefont {L.}~\bibnamefont {Ge}}, \bibinfo {author} {\bibfnamefont
  {M.}~\bibnamefont {Steger}}, \bibinfo {author} {\bibfnamefont {L.~N.}\
  \bibnamefont {Pfeiffer}}, \bibinfo {author} {\bibfnamefont {K.}~\bibnamefont
  {West}}, \bibinfo {author} {\bibfnamefont {H.~E.}\ \bibnamefont {T\"ureci}},
  \bibinfo {author} {\bibfnamefont {D.~W.}\ \bibnamefont {Snoke}},\ and\
  \bibinfo {author} {\bibfnamefont {K.~A.}\ \bibnamefont {Nelson}},\ }\bibfield
   {title} {\bibinfo {title} {Stable switching among high-order modes in
  polariton condensates},\ }\href {https://doi.org/10.1103/PhysRevB.97.045303}
  {\bibfield  {journal} {\bibinfo  {journal} {Phys. Rev. B}\ }\textbf {\bibinfo
  {volume} {97}},\ \bibinfo {pages} {045303} (\bibinfo {year}
  {2018})}\BibitemShut {NoStop}%
\bibitem [{\citenamefont {Xu}\ \emph {et~al.}(2019)\citenamefont {Xu},
  \citenamefont {Zhou}, \citenamefont {Wang}, \citenamefont {Tian},
  \citenamefont {Zhang}, \citenamefont {Dong}, \citenamefont {Wang},\ and\
  \citenamefont {Zhou}}]{Xu_OptExp2019}%
  \BibitemOpen
  \bibfield  {author} {\bibinfo {author} {\bibfnamefont {C.}~\bibnamefont
  {Xu}}, \bibinfo {author} {\bibfnamefont {B.}~\bibnamefont {Zhou}}, \bibinfo
  {author} {\bibfnamefont {X.}~\bibnamefont {Wang}}, \bibinfo {author}
  {\bibfnamefont {C.}~\bibnamefont {Tian}}, \bibinfo {author} {\bibfnamefont
  {Y.}~\bibnamefont {Zhang}}, \bibinfo {author} {\bibfnamefont
  {H.}~\bibnamefont {Dong}}, \bibinfo {author} {\bibfnamefont {G.}~\bibnamefont
  {Wang}},\ and\ \bibinfo {author} {\bibfnamefont {W.}~\bibnamefont {Zhou}},\
  }\bibfield  {title} {\bibinfo {title} {Dynamics of excited-state condensate
  for optically confined exciton-polaritons},\ }\href
  {https://doi.org/10.1364/OE.27.024938} {\bibfield  {journal} {\bibinfo
  {journal} {Opt. Express}\ }\textbf {\bibinfo {volume} {27}},\ \bibinfo
  {pages} {24938} (\bibinfo {year} {2019})}\BibitemShut {NoStop}%
\bibitem [{\citenamefont {Berloff}\ \emph {et~al.}(2017)\citenamefont
  {Berloff}, \citenamefont {Silva}, \citenamefont {Kalinin}, \citenamefont
  {Askitopoulos}, \citenamefont {T{\"o}pfer}, \citenamefont {Cilibrizzi},
  \citenamefont {Langbein},\ and\ \citenamefont
  {Lagoudakis}}]{berloff_realizing_2016}%
  \BibitemOpen
  \bibfield  {author} {\bibinfo {author} {\bibfnamefont {N.~G.}\ \bibnamefont
  {Berloff}}, \bibinfo {author} {\bibfnamefont {M.}~\bibnamefont {Silva}},
  \bibinfo {author} {\bibfnamefont {K.}~\bibnamefont {Kalinin}}, \bibinfo
  {author} {\bibfnamefont {A.}~\bibnamefont {Askitopoulos}}, \bibinfo {author}
  {\bibfnamefont {J.~D.}\ \bibnamefont {T{\"o}pfer}}, \bibinfo {author}
  {\bibfnamefont {P.}~\bibnamefont {Cilibrizzi}}, \bibinfo {author}
  {\bibfnamefont {W.}~\bibnamefont {Langbein}},\ and\ \bibinfo {author}
  {\bibfnamefont {P.~G.}\ \bibnamefont {Lagoudakis}},\ }\bibfield  {title}
  {\bibinfo {title} {Realizing the classical xy hamiltonian in polariton
  simulators},\ }\href {https://doi.org/10.1038/nmat4971} {\bibfield  {journal}
  {\bibinfo  {journal} {Nature Materials}\ }\textbf {\bibinfo {volume} {16}},\
  \bibinfo {pages} {1120} (\bibinfo {year} {2017})}\BibitemShut {NoStop}%
\bibitem [{\citenamefont {Johnston}\ \emph {et~al.}(2020)\citenamefont
  {Johnston}, \citenamefont {Kalinin},\ and\ \citenamefont
  {Berloff}}]{alex2020artificial}%
  \BibitemOpen
  \bibfield  {author} {\bibinfo {author} {\bibfnamefont {A.}~\bibnamefont
  {Johnston}}, \bibinfo {author} {\bibfnamefont {K.~P.}\ \bibnamefont
  {Kalinin}},\ and\ \bibinfo {author} {\bibfnamefont {N.~G.}\ \bibnamefont
  {Berloff}},\ }\href@noop {} {\bibinfo {title} {Artificial polariton
  molecules}} (\bibinfo {year} {2020}),\ \Eprint
  {https://arxiv.org/abs/2006.10492} {arXiv:2006.10492 [cond-mat.mes-hall]}
  \BibitemShut {NoStop}%
\bibitem [{\citenamefont {Wouters}\ and\ \citenamefont
  {Carusotto}(2007)}]{Wouters2007a}%
  \BibitemOpen
  \bibfield  {author} {\bibinfo {author} {\bibfnamefont {M.}~\bibnamefont
  {Wouters}}\ and\ \bibinfo {author} {\bibfnamefont {I.}~\bibnamefont
  {Carusotto}},\ }\bibfield  {title} {\bibinfo {title} {Excitations in a
  nonequilibrium bose-einstein condensate of exciton polaritons},\ }\href
  {https://doi.org/10.1103/PhysRevLett.99.140402} {\bibfield  {journal}
  {\bibinfo  {journal} {Phys. Rev. Lett.}\ }\textbf {\bibinfo {volume} {99}},\
  \bibinfo {pages} {140402} (\bibinfo {year} {2007})}\BibitemShut {NoStop}%
\bibitem [{\citenamefont {Harrison}\ \emph {et~al.}(2020)\citenamefont
  {Harrison}, \citenamefont {Sigurdsson},\ and\ \citenamefont
  {Lagoudakis}}]{Harrison_PRB2020}%
  \BibitemOpen
  \bibfield  {author} {\bibinfo {author} {\bibfnamefont {S.~L.}\ \bibnamefont
  {Harrison}}, \bibinfo {author} {\bibfnamefont {H.}~\bibnamefont
  {Sigurdsson}},\ and\ \bibinfo {author} {\bibfnamefont {P.~G.}\ \bibnamefont
  {Lagoudakis}},\ }\bibfield  {title} {\bibinfo {title} {Synchronization in
  optically trapped polariton stuart-landau networks},\ }\href
  {https://doi.org/10.1103/PhysRevB.101.155402} {\bibfield  {journal} {\bibinfo
   {journal} {Phys. Rev. B}\ }\textbf {\bibinfo {volume} {101}},\ \bibinfo
  {pages} {155402} (\bibinfo {year} {2020})}\BibitemShut {NoStop}%
\bibitem [{\citenamefont {Aleiner}\ \emph {et~al.}(2012)\citenamefont
  {Aleiner}, \citenamefont {Altshuler},\ and\ \citenamefont
  {Rubo}}]{Aleiner2012}%
  \BibitemOpen
  \bibfield  {author} {\bibinfo {author} {\bibfnamefont {I.~L.}\ \bibnamefont
  {Aleiner}}, \bibinfo {author} {\bibfnamefont {B.~L.}\ \bibnamefont
  {Altshuler}},\ and\ \bibinfo {author} {\bibfnamefont {Y.~G.}\ \bibnamefont
  {Rubo}},\ }\bibfield  {title} {\bibinfo {title} {Radiative coupling and weak
  lasing of exciton-polariton condensates},\ }\href
  {https://doi.org/10.1103/PhysRevB.85.121301} {\bibfield  {journal} {\bibinfo
  {journal} {Phys. Rev. B}\ }\textbf {\bibinfo {volume} {85}},\ \bibinfo
  {pages} {121301} (\bibinfo {year} {2012})}\BibitemShut {NoStop}%
\bibitem [{Note1()}]{Note1}%
  \BibitemOpen
  \bibinfo {note} {Condensation into the state $n=1$ was observed in Ref.~\cite
  {Dreismann_PNAS2014} when the pumping had the shape of two concentric
  rings}\BibitemShut {NoStop}%
\bibitem [{\citenamefont {Ohadi}\ \emph {et~al.}(2016)\citenamefont {Ohadi},
  \citenamefont {Gregory}, \citenamefont {Freegarde}, \citenamefont {Rubo},
  \citenamefont {Kavokin}, \citenamefont {Berloff},\ and\ \citenamefont
  {Lagoudakis}}]{Ohadi_PRX2016}%
  \BibitemOpen
  \bibfield  {author} {\bibinfo {author} {\bibfnamefont {H.}~\bibnamefont
  {Ohadi}}, \bibinfo {author} {\bibfnamefont {R.~L.}\ \bibnamefont {Gregory}},
  \bibinfo {author} {\bibfnamefont {T.}~\bibnamefont {Freegarde}}, \bibinfo
  {author} {\bibfnamefont {Y.~G.}\ \bibnamefont {Rubo}}, \bibinfo {author}
  {\bibfnamefont {A.~V.}\ \bibnamefont {Kavokin}}, \bibinfo {author}
  {\bibfnamefont {N.~G.}\ \bibnamefont {Berloff}},\ and\ \bibinfo {author}
  {\bibfnamefont {P.~G.}\ \bibnamefont {Lagoudakis}},\ }\bibfield  {title}
  {\bibinfo {title} {Nontrivial phase coupling in polariton multiplets},\
  }\href {https://doi.org/10.1103/PhysRevX.6.031032} {\bibfield  {journal}
  {\bibinfo  {journal} {Phys. Rev. X}\ }\textbf {\bibinfo {volume} {6}},\
  \bibinfo {pages} {031032} (\bibinfo {year} {2016})}\BibitemShut {NoStop}%
\bibitem [{\citenamefont {Zhang}\ \emph {et~al.}(2015)\citenamefont {Zhang},
  \citenamefont {Xie}, \citenamefont {Wang}, \citenamefont {Poddubny},
  \citenamefont {Lu}, \citenamefont {Wang}, \citenamefont {Gu}, \citenamefont
  {Liu}, \citenamefont {Xu}, \citenamefont {Shen}, \citenamefont {Rubo},
  \citenamefont {Altshuler}, \citenamefont {Kavokin},\ and\ \citenamefont
  {Chen}}]{Zhang2015}%
  \BibitemOpen
  \bibfield  {author} {\bibinfo {author} {\bibfnamefont {L.}~\bibnamefont
  {Zhang}}, \bibinfo {author} {\bibfnamefont {W.}~\bibnamefont {Xie}}, \bibinfo
  {author} {\bibfnamefont {J.}~\bibnamefont {Wang}}, \bibinfo {author}
  {\bibfnamefont {A.}~\bibnamefont {Poddubny}}, \bibinfo {author}
  {\bibfnamefont {J.}~\bibnamefont {Lu}}, \bibinfo {author} {\bibfnamefont
  {Y.}~\bibnamefont {Wang}}, \bibinfo {author} {\bibfnamefont {J.}~\bibnamefont
  {Gu}}, \bibinfo {author} {\bibfnamefont {W.}~\bibnamefont {Liu}}, \bibinfo
  {author} {\bibfnamefont {D.}~\bibnamefont {Xu}}, \bibinfo {author}
  {\bibfnamefont {X.}~\bibnamefont {Shen}}, \bibinfo {author} {\bibfnamefont
  {Y.~G.}\ \bibnamefont {Rubo}}, \bibinfo {author} {\bibfnamefont {B.~L.}\
  \bibnamefont {Altshuler}}, \bibinfo {author} {\bibfnamefont {A.~V.}\
  \bibnamefont {Kavokin}},\ and\ \bibinfo {author} {\bibfnamefont
  {Z.}~\bibnamefont {Chen}},\ }\bibfield  {title} {\bibinfo {title} {{Weak
  lasing in one-dimensional polariton superlattices}},\ }\href
  {https://doi.org/10.1073/pnas.1502666112} {\bibfield  {journal} {\bibinfo
  {journal} {Proceedings of the National Academy of Sciences}\ }\textbf
  {\bibinfo {volume} {112}},\ \bibinfo {pages} {E1516} (\bibinfo {year}
  {2015})}\BibitemShut {NoStop}%
\bibitem [{\citenamefont {Nalitov}\ \emph {et~al.}(2019)\citenamefont
  {Nalitov}, \citenamefont {Sigurdsson}, \citenamefont {Morina}, \citenamefont
  {Krivosenko}, \citenamefont {Iorsh}, \citenamefont {Rubo}, \citenamefont
  {Kavokin},\ and\ \citenamefont {Shelykh}}]{Nalitov_PRA2019}%
  \BibitemOpen
  \bibfield  {author} {\bibinfo {author} {\bibfnamefont {A.~V.}\ \bibnamefont
  {Nalitov}}, \bibinfo {author} {\bibfnamefont {H.}~\bibnamefont {Sigurdsson}},
  \bibinfo {author} {\bibfnamefont {S.}~\bibnamefont {Morina}}, \bibinfo
  {author} {\bibfnamefont {Y.~S.}\ \bibnamefont {Krivosenko}}, \bibinfo
  {author} {\bibfnamefont {I.~V.}\ \bibnamefont {Iorsh}}, \bibinfo {author}
  {\bibfnamefont {Y.~G.}\ \bibnamefont {Rubo}}, \bibinfo {author}
  {\bibfnamefont {A.~V.}\ \bibnamefont {Kavokin}},\ and\ \bibinfo {author}
  {\bibfnamefont {I.~A.}\ \bibnamefont {Shelykh}},\ }\bibfield  {title}
  {\bibinfo {title} {Optically trapped polariton condensates as semiclassical
  time crystals},\ }\href {https://doi.org/10.1103/PhysRevA.99.033830}
  {\bibfield  {journal} {\bibinfo  {journal} {Phys. Rev. A}\ }\textbf {\bibinfo
  {volume} {99}},\ \bibinfo {pages} {033830} (\bibinfo {year}
  {2019})}\BibitemShut {NoStop}%
\bibitem [{\citenamefont {T{\"o}pfer}\ \emph {et~al.}(2020)\citenamefont
  {T{\"o}pfer}, \citenamefont {Sigurdsson}, \citenamefont {Pickup},\ and\
  \citenamefont {Lagoudakis}}]{Topfer_ComPhys2020}%
  \BibitemOpen
  \bibfield  {author} {\bibinfo {author} {\bibfnamefont {J.~D.}\ \bibnamefont
  {T{\"o}pfer}}, \bibinfo {author} {\bibfnamefont {H.}~\bibnamefont
  {Sigurdsson}}, \bibinfo {author} {\bibfnamefont {L.}~\bibnamefont {Pickup}},\
  and\ \bibinfo {author} {\bibfnamefont {P.~G.}\ \bibnamefont {Lagoudakis}},\
  }\bibfield  {title} {\bibinfo {title} {Time-delay polaritonics},\ }\href
  {https://doi.org/10.1038/s42005-019-0271-0} {\bibfield  {journal} {\bibinfo
  {journal} {Communications Physics}\ }\textbf {\bibinfo {volume} {3}},\
  \bibinfo {pages} {2} (\bibinfo {year} {2020})}\BibitemShut {NoStop}%
\bibitem [{\citenamefont {Ruiz-S\'anchez}\ \emph {et~al.}(2020)\citenamefont
  {Ruiz-S\'anchez}, \citenamefont {Rechtman},\ and\ \citenamefont
  {Rubo}}]{Ruiz_PRB2020}%
  \BibitemOpen
  \bibfield  {author} {\bibinfo {author} {\bibfnamefont {R.}~\bibnamefont
  {Ruiz-S\'anchez}}, \bibinfo {author} {\bibfnamefont {R.}~\bibnamefont
  {Rechtman}},\ and\ \bibinfo {author} {\bibfnamefont {Y.~G.}\ \bibnamefont
  {Rubo}},\ }\bibfield  {title} {\bibinfo {title} {Autonomous chaos of
  exciton-polariton condensates},\ }\href
  {https://doi.org/10.1103/PhysRevB.101.155305} {\bibfield  {journal} {\bibinfo
   {journal} {Phys. Rev. B}\ }\textbf {\bibinfo {volume} {101}},\ \bibinfo
  {pages} {155305} (\bibinfo {year} {2020})}\BibitemShut {NoStop}%
\bibitem [{\citenamefont {Cerna}\ \emph {et~al.}(2009)\citenamefont {Cerna},
  \citenamefont {Sarchi}, \citenamefont {Para\"{\i}so}, \citenamefont {Nardin},
  \citenamefont {L\'eger}, \citenamefont {Richard}, \citenamefont {Pietka},
  \citenamefont {El~Daif}, \citenamefont {Morier-Genoud}, \citenamefont
  {Savona}, \citenamefont {Portella-Oberli},\ and\ \citenamefont
  {Deveaud-Pl\'edran}}]{Cerna_PRB2009}%
  \BibitemOpen
  \bibfield  {author} {\bibinfo {author} {\bibfnamefont {R.}~\bibnamefont
  {Cerna}}, \bibinfo {author} {\bibfnamefont {D.}~\bibnamefont {Sarchi}},
  \bibinfo {author} {\bibfnamefont {T.~K.}\ \bibnamefont {Para\"{\i}so}},
  \bibinfo {author} {\bibfnamefont {G.}~\bibnamefont {Nardin}}, \bibinfo
  {author} {\bibfnamefont {Y.}~\bibnamefont {L\'eger}}, \bibinfo {author}
  {\bibfnamefont {M.}~\bibnamefont {Richard}}, \bibinfo {author} {\bibfnamefont
  {B.}~\bibnamefont {Pietka}}, \bibinfo {author} {\bibfnamefont
  {O.}~\bibnamefont {El~Daif}}, \bibinfo {author} {\bibfnamefont
  {F.}~\bibnamefont {Morier-Genoud}}, \bibinfo {author} {\bibfnamefont
  {V.}~\bibnamefont {Savona}}, \bibinfo {author} {\bibfnamefont {M.~T.}\
  \bibnamefont {Portella-Oberli}},\ and\ \bibinfo {author} {\bibfnamefont
  {B.}~\bibnamefont {Deveaud-Pl\'edran}},\ }\bibfield  {title} {\bibinfo
  {title} {Coherent optical control of the wave function of zero-dimensional
  exciton polaritons},\ }\href {https://doi.org/10.1103/PhysRevB.80.121309}
  {\bibfield  {journal} {\bibinfo  {journal} {Phys. Rev. B}\ }\textbf {\bibinfo
  {volume} {80}},\ \bibinfo {pages} {121309} (\bibinfo {year}
  {2009})}\BibitemShut {NoStop}%
\bibitem [{\citenamefont {Askitopoulos}\ \emph {et~al.}(2018)\citenamefont
  {Askitopoulos}, \citenamefont {Nalitov}, \citenamefont {Sedov}, \citenamefont
  {Pickup}, \citenamefont {Cherotchenko}, \citenamefont {Hatzopoulos},
  \citenamefont {Savvidis}, \citenamefont {Kavokin},\ and\ \citenamefont
  {Lagoudakis}}]{askitopoulos_all-optical_2018}%
  \BibitemOpen
  \bibfield  {author} {\bibinfo {author} {\bibfnamefont {A.}~\bibnamefont
  {Askitopoulos}}, \bibinfo {author} {\bibfnamefont {A.~V.}\ \bibnamefont
  {Nalitov}}, \bibinfo {author} {\bibfnamefont {E.~S.}\ \bibnamefont {Sedov}},
  \bibinfo {author} {\bibfnamefont {L.}~\bibnamefont {Pickup}}, \bibinfo
  {author} {\bibfnamefont {E.~D.}\ \bibnamefont {Cherotchenko}}, \bibinfo
  {author} {\bibfnamefont {Z.}~\bibnamefont {Hatzopoulos}}, \bibinfo {author}
  {\bibfnamefont {P.~G.}\ \bibnamefont {Savvidis}}, \bibinfo {author}
  {\bibfnamefont {A.~V.}\ \bibnamefont {Kavokin}},\ and\ \bibinfo {author}
  {\bibfnamefont {P.~G.}\ \bibnamefont {Lagoudakis}},\ }\bibfield  {title}
  {\bibinfo {title} {All-optical quantum fluid spin beam splitter},\ }\href
  {https://doi.org/10.1103/PhysRevB.97.235303} {\bibfield  {journal} {\bibinfo
  {journal} {Phys. Rev. B}\ }\textbf {\bibinfo {volume} {97}},\ \bibinfo
  {pages} {235303} (\bibinfo {year} {2018})}\BibitemShut {NoStop}%
\end{thebibliography}%

\end{document}